\documentclass[a4paper,fleqn,usenatbib]{mnras}

\usepackage[T1]{fontenc}
\usepackage{ae,aecompl}

\usepackage{graphicx}	
\usepackage{amsmath}	
\usepackage{amssymb}	
\usepackage{xfrac}
\usepackage{makecell}
\usepackage{multirow}
\usepackage{ulem}

\newcommand{\kms}{km\,s$^{-1}$}

\title[On the dust and gas components of RX J0911]
      {On the dust and gas components of the $z=2.8$ gravitationally lensed quasar host RX J0911.4+0551}
\author[P. Tuan-Anh et al.]{
{P. Tuan-Anh\thanks{E-mail: ptanh@vnsc.org.vn}, D. T. Hoai, P. T. Nhung, P. N. Diep, N. T. Phuong, N. T. Thao}
\newauthor{ and P. Darriulat}
\\
Department of Astrophysics, Vietnam National Satellite Center, VAST, 18 Hoang Quoc Viet, Hanoi, Vietnam\\
}
\date{Accepted XXX. Received YYY; in original form ZZZ}

\pubyear{2016}

\begin{document}
\label{firstpage}
\pagerange{\pageref{firstpage}--\pageref{lastpage}}
\maketitle

\begin{abstract}
  Observations by the Atacama Large Millimetre/sub-millimetre Array of the 358 GHz continuum emission of the gravitationally lensed quasar host RX J0911.4+0551 have been analysed. They complement earlier Plateau de Bure Interferometer observations of the CO(7-6) emission. The good knowledge of the lensing potential obtained from Hubble Space Telescope observations of the quasar makes a joint analysis of the three emissions possible. It gives evidence for the quasar source to be concentric with the continuum source within 0.31 kpc and with the CO(7-6) source within 1.10 kpc. It also provides a measurement of the size of the continuum source, 0.76 $\pm$ 0.04 kpc FWHM, making RX J0911.4+0551 one of the few high redshift galaxies for which the dust and gas components are resolved with dimensions being measured. Both are found to be very compact, the former being smaller than the latter by a factor of $\sim$3.4$\pm$0.4. Moreover, new measurements of the CO ladder $-$ CO(10-9) and CO(11-10) $-$ are presented that confirm the extreme narrowness of the CO line width (107$\pm$20 km s$^{-1}$ on average). Their mere detection implies higher temperature and/or density than for typical quasar hosts at this redshift and suggests a possible contribution of the central AGN to gas and dust heating. The results are interpreted in terms of current understanding of galaxy evolution at the peak of star formation. They suggest that RX J0911.4+0551 is a young galaxy in an early stage of its evolution, having experienced no recent major mergers, star formation being concentrated in its centre. 
\end{abstract}




\section{Introduction}
Our knowledge of galaxy evolution at early cosmic times, in particular at the epoch when star formation and AGN activity reach their maximum, with redshift in the $\sim$2 to $\sim$4.5 range, owes much to the study of quasar hosts (for a review see \citealt{Carilli2013}, for recent new results see \citealt{Aravena2016,Glikman2016}). Four main actors of this evolution are accessible to observation: the AGN emission around the supermassive black hole in the centre, the gas reservoir from which stars are formed, its dust content $-$ the emission of which is a proxy of star formation $-$ and the stars themselves. The fifth main actor, dark matter, is not directly accessible to observation. Typically, each of these covers a specific frequency range in the rest frame: optical and X rays for the black hole, millimetre/sub-millimetre for the molecular gas, infrared for the dust and optical for the stars. These observations are reduced to a few quantities that summarize our knowledge of the observed galaxy, such as masses and luminosities of the above mentioned components, star formation rate, dynamical mass, starburstiness, etc.

The molecule that is most commonly used as a tracer of the gas component is carbon monoxide; the emission of the molecular rotation levels probes different regions depending on the excitation energy. Other tracers, such as carbon and nitrogen fine structure lines, have also been detected successfully. It is only recently that some gas and dust components of high redshift quasar hosts and sub-millimetre galaxies could be spatially resolved, usually taking advantage of the important magnification provided by gravitational lensing \citep{Riechers2011a,Riechers2011b,Riechers2011c}. One of these is the host galaxy of RX J0911.4+0551 (hereafter referred to as RX J0911), which we studied earlier using Plateau de Bure Interferometer observations \citep{Tuananh2013,Tuananh2014,Hoai2013} and which is the subject of the present article.

While many quasar hosts and sub-millimetre galaxies hosting an AGN have been detected at redshifts between $\sim$2 and $\sim$4.5, most of them have been observed in the continuum and only few in molecular or atomic lines. Table \ref{Table1} lists some typical cases chosen among the better resolved. In the case of gravitational lensing it is not sufficient to resolve the lensed images with good spatial resolution; in addition, a precise knowledge of the lensing potential is mandatory in order to resolve the source and measure its size. The resolved images, while giving often evidence for extended dust and gas components, the former being relatively more compact, benefit rarely from a sufficient spatial resolution for their morphology to be reconstructed in the source plane. 

From the study of high redshift galaxies, one learns that they are often the seat of mergers, in particular wet mergers that are identified by comparing the respective locations of the optical, gas and dust components. Examples are J 123707, SMM J02399 and BRI 1335. Typically, mergers are identified with dimensions in excess of a kiloparsec in the source plane. Mergers cause the gravitational field to strongly increase locally, in particular as the result of induced shocks and turbulences, triggering the local collapse of gas clouds and causing star bursts. These are seen as important sources of dust away from the central supermassive black hole. In cases where the gas and dust components are roughly concentric with the supermassive black hole, with dimensions not exceeding a kiloparsec in the source plane, a velocity gradient is often observed that reveals a disc-like rotating structure; in most cases these give no evidence for recent mergers but the better resolution observations may display a clumpy structure of the source, possibly revealing the effect of merging: such is the case of SDP.81, the best resolved of all cases.

In summary, understanding the genesis of early galaxies requires observations made at different stages of their evolution, in order to reveal the relative roles played by each actor as a function of time. At each stage, multi-wavelength observations are mandatory in order to disentangle the respective morphologies of the gas and dust components and their locations with respect to the central supermassive black hole.

The next section, Section 2, recalls what is known of RX J0911; Section 3 introduces the new observations and the reduction of the data, with Section 4 discussing the relative positions of the quasar and the gas and dust components and Section 5 discussing their sizes. Section 6
locates the continuum flux measurement on the SED and Section 7 discusses the line data. The article closes with a summary of the main results and their interpretation in the framework of our present knowledge of galaxy evolution at high redshifts.
\begin{table*}
\centering 
\caption{Some typical quasar hosts and sub-millimetre galaxies at redshifts in the 2.5 to 4.5 range for which both the gas and dust components have been resolved and measured. References are given for the line emission data, the continuum emission data and the lensing mechanism. In the last column, mergers are indicated as M and others as NM; in both cases, the figure gives the scale of the emission in kpc.}
\label{Table1}        
\begin{tabular}{|c|c|c|c|c|c|}
\hline 
Name & $z$ & Ref. line & Ref. continuum & Ref. lens & Mergers \\
\hline 
J123707+6214 & 2.5 & CO(1-0) \& (5-4)[20] & 1.4 GHz, [20,21] & unlensed & M, 20 \\
\hline
Cloverleaf & 2.56 & CO(7-6) [6] & 122 microns [7] & [6] & NM, 0.8 \\
\hline
SPT0538-50 & 2.78 & CO(1-0) \&(3-2) [8] & 860 microns [9] & [8,9] & M, 1.6 \\
\hline
RX J0911 & 2.8 & CO(7-6) [1,2,4] & [5] & [3,4] & NM, 0.8 \\
\hline
SMMJ02399-0136 & 2.81 & CO(1-0) [10] & 122 microns [7] & weakly lensed [11] & M, 25 \\
\hline
SDP.81 & 3.04 &  CO(5-4) \& (8-7) [12,14] & 236 and 290 GHz [12,13] & [13,14] & M, 8\\
\hline
APM 08279+5255 & 3.91 & CO(1-0) [15] & 2.6 mm [15] & [15] & NM, 0.5 \\
\hline
PSS J2322+1944 & 4.12 & CO(2-1) [16] & 1.4 GHz [17] & [16,17] & NM, 2 \\
\hline
BRI 1335-0417 & 4.41 & CO(2-1) [18] & 1.4 GHz [19] & unlensed & M, 5 \\
\hline

\end{tabular} 

1. \citet{Weiss2012}; 2. \citet{Tuananh2013}; 3. \citet{Hoai2013}; 4. \citet{Tuananh2014};
5. ALMA archive, this work; \\6. \citet{Venturini2003}; 7. \citet{Ferkinhoff2015};
8. \citet{Spilker2015}; 9. \citet{Hezaveh2013}; 10. \citet{Ivison2010};\\
11. \citet{Richard2009}; 12. \citet{ALMA2015}; 13. \citet{Rybak2015a};
14. \citet{Rybak2015b};\\ 15. \citet{Riechers2009a}; 16. \citet{Riechers2008a};
17. \citet{Carilli2001} \& \citet{Carilli2003}; 18. \citet{Riechers2008b}; 19. \citet{Momjian2007};
20. \citet{Riechers2011e}; 21. \citet{Morrison2010}.\\
\end{table*}

\section{RX J0911 and its host galaxy}
\begin{figure*}
\centering
\includegraphics[height=4.5cm,trim=-3.cm -2.2cm -.5cm 0.cm,clip]{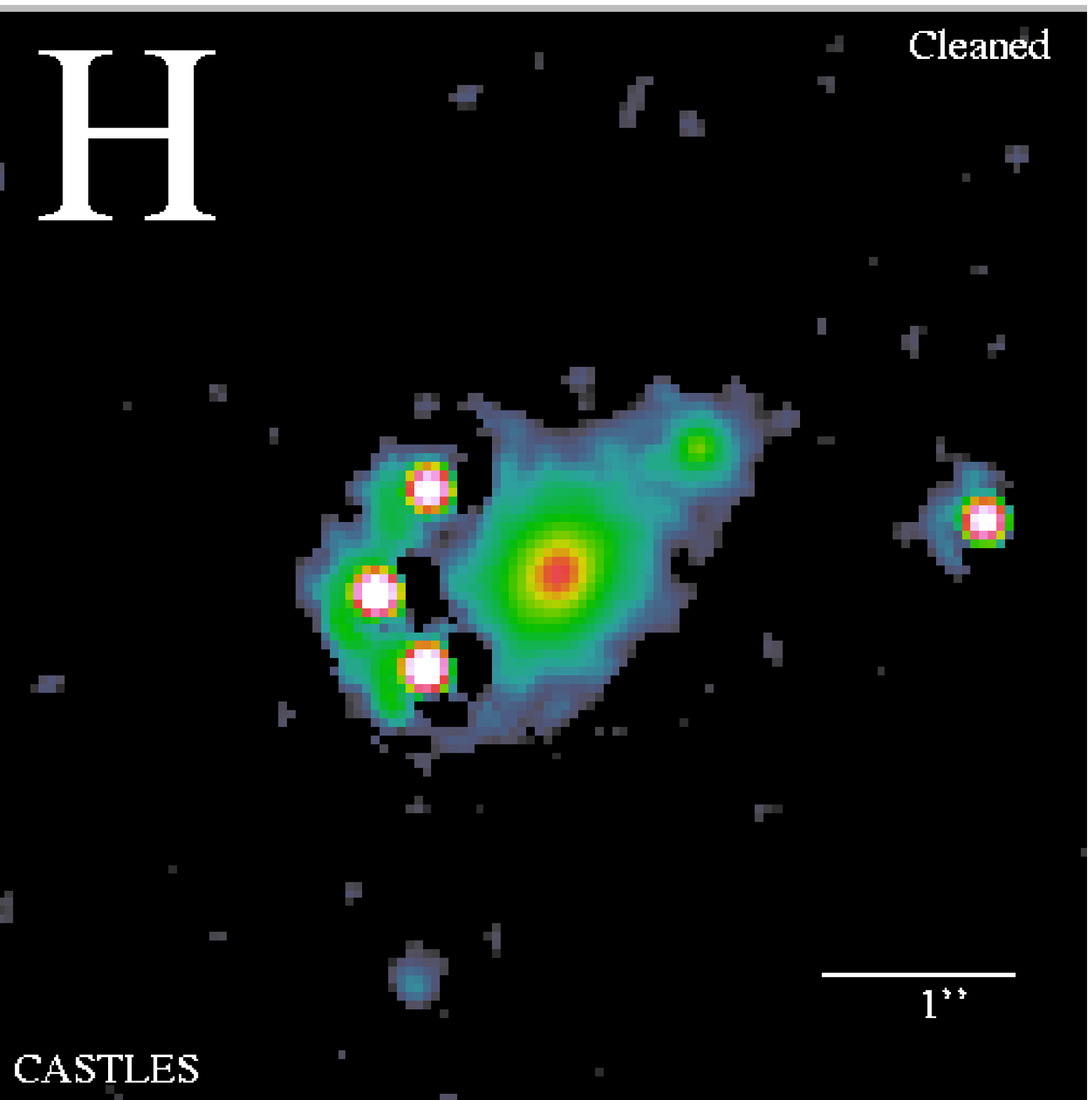}
\includegraphics[height=5.cm,trim=0.cm 0.cm 0.5cm 0.cm,clip]{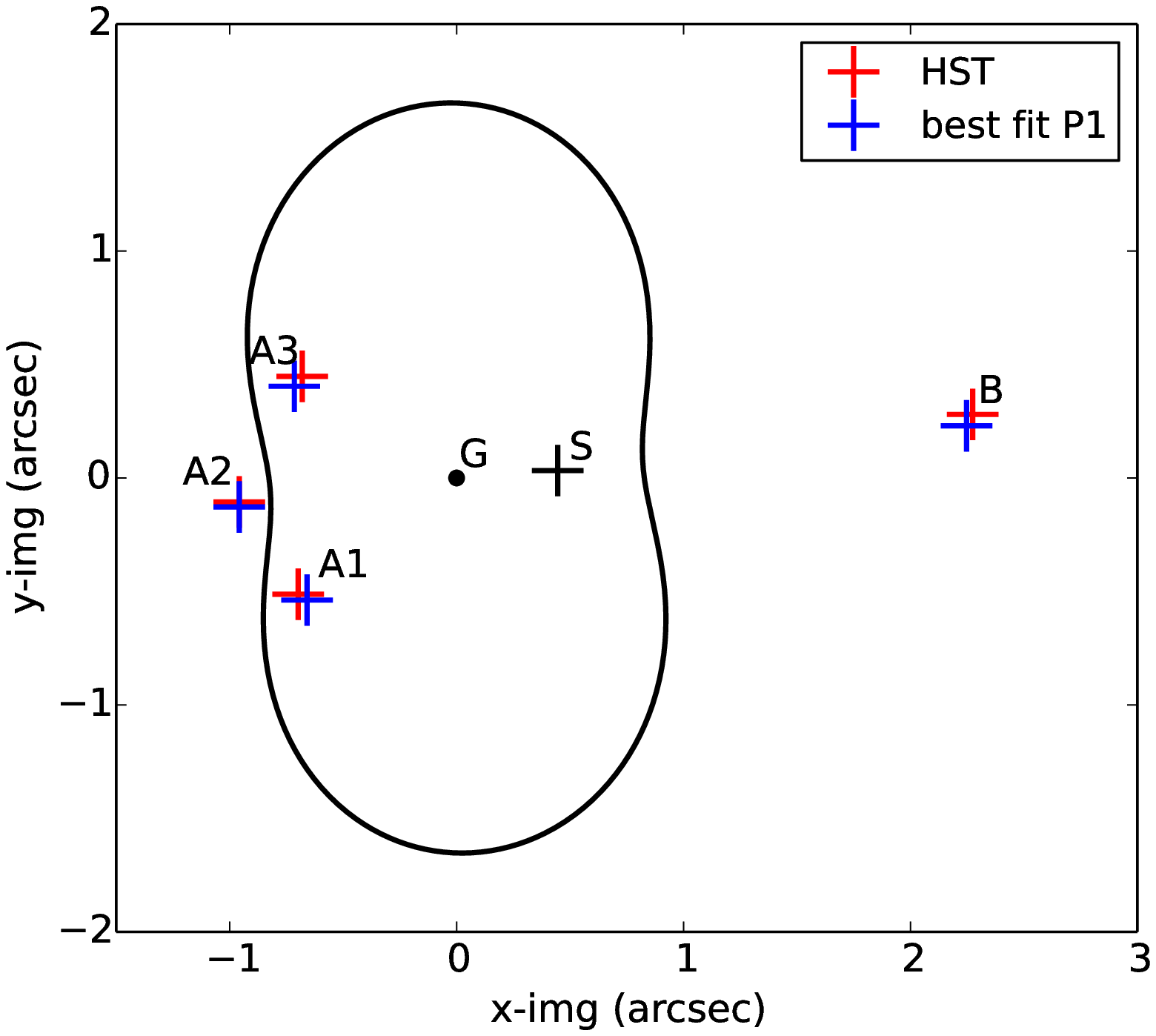}
\includegraphics[height=5.cm,trim=2.4cm 0.cm 0.cm 0.cm,clip]{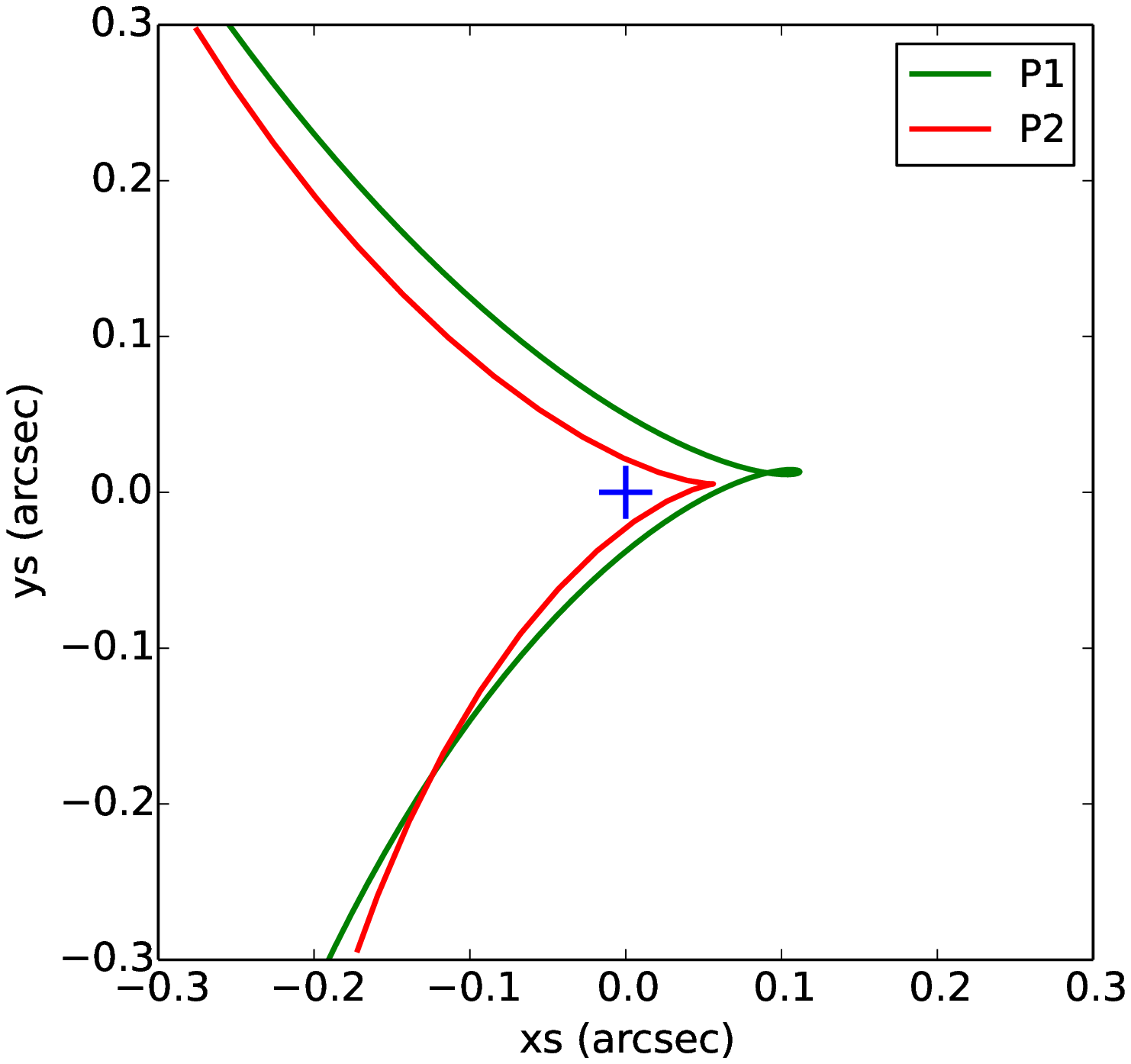}
\caption{RX J0911. Left: HST/Castles images of the quasar. Middle: Positions of the lensed quasar images observed by HST (red crosses) and obtained from the lensing potential (blue crosses) are shown together with the critical curve (the main lens galaxy G is at the origin of coordinates, S shows the source position). Right: Caustic curves in the source plane for both lensing potentials, the source is at the origin of coordinates, marked by a cross.}
\label{fig1}
\end{figure*}

RX J0911 was first observed in 1995 in X-ray by ROSAT \citep{Bade1995,Bade1997,Hagen1995}, then in the optical and near-infrared using NOT (Nordic Optical Telescope) and NTT (New Technology Telescope) by \citet{Burud1998a,Burud1998b} and later (Figure \ref{fig1} left) using the Hubble Space Telescope (HST) by \citet{Kochanek2002}. It is a $z=2.8$ quasar. Using $\Lambda$CDM concordance cosmology (\mbox{H$_0$=71 \kms Mpc$^{-1}$}, $\Omega_M$=0.27 and $\Omega_\Lambda$=0.73), this implies that an angular distance of 1 arcsec corresponds to a real distance of 7.98 kpc. The quasar is lensed into four resolved images, typical of a source located near a cusp of the minor axis of the caustic \citep{Saha2003}: a triplet of nearby A images in the east and a fainter B image in the west, separated from the A images by some 3 arcsec (Figure \ref{fig1} middle and right). The lens has been identified as a galaxy G at $z=0.77$, with the cluster to which it belongs \citep{Kneib2000} causing a small shear. The time delay between the A and B images ($\sim$150 days) has been measured by \citet{Hjorth2002}.

\begin{figure*}
  \centering
  \includegraphics[height=4.25cm,trim=0.cm 0.cm 0.cm 0.cm,clip]{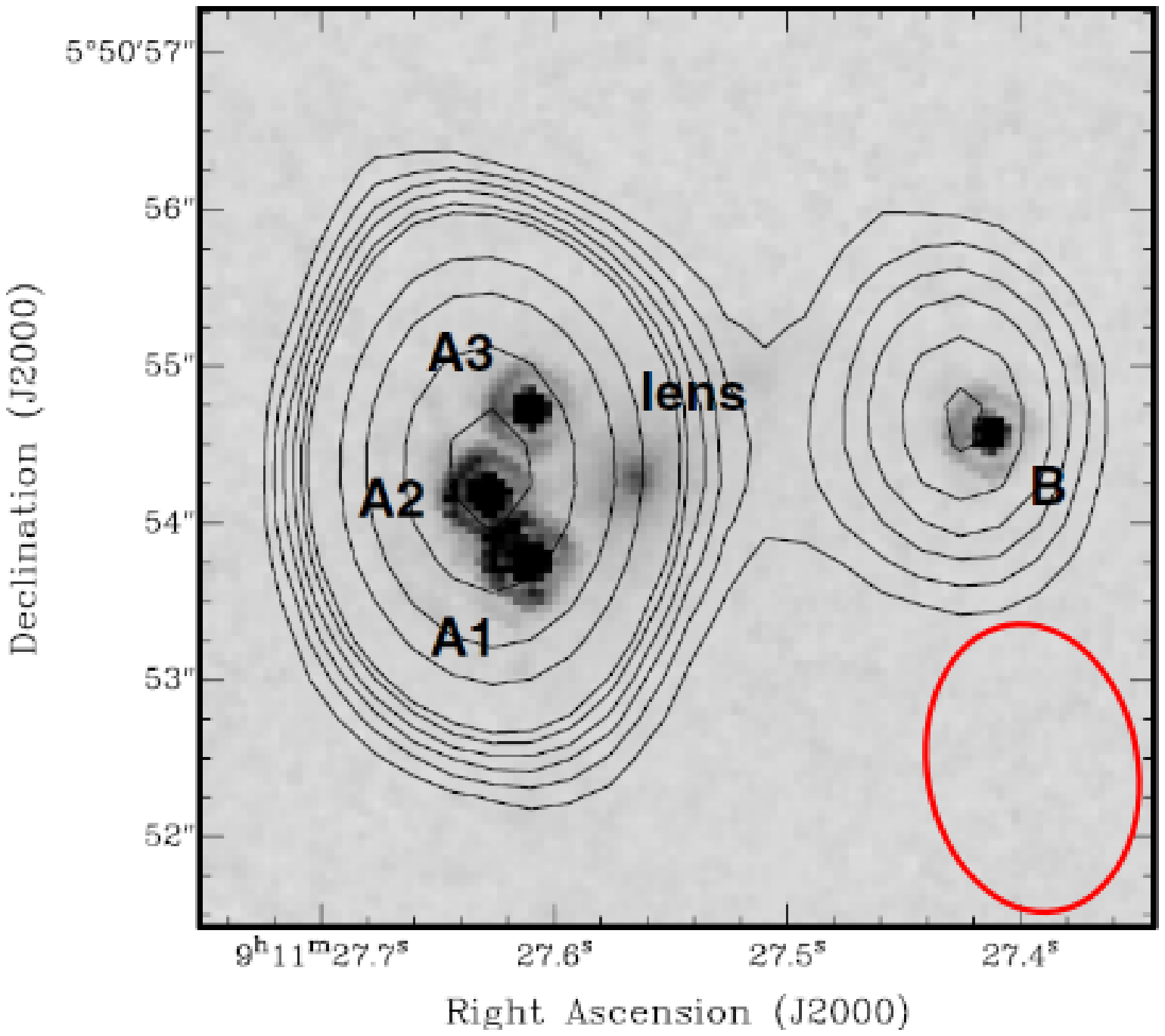}
  \includegraphics[height=4.5cm,trim=-2.5cm 0.cm 0.cm 0.cm,clip]{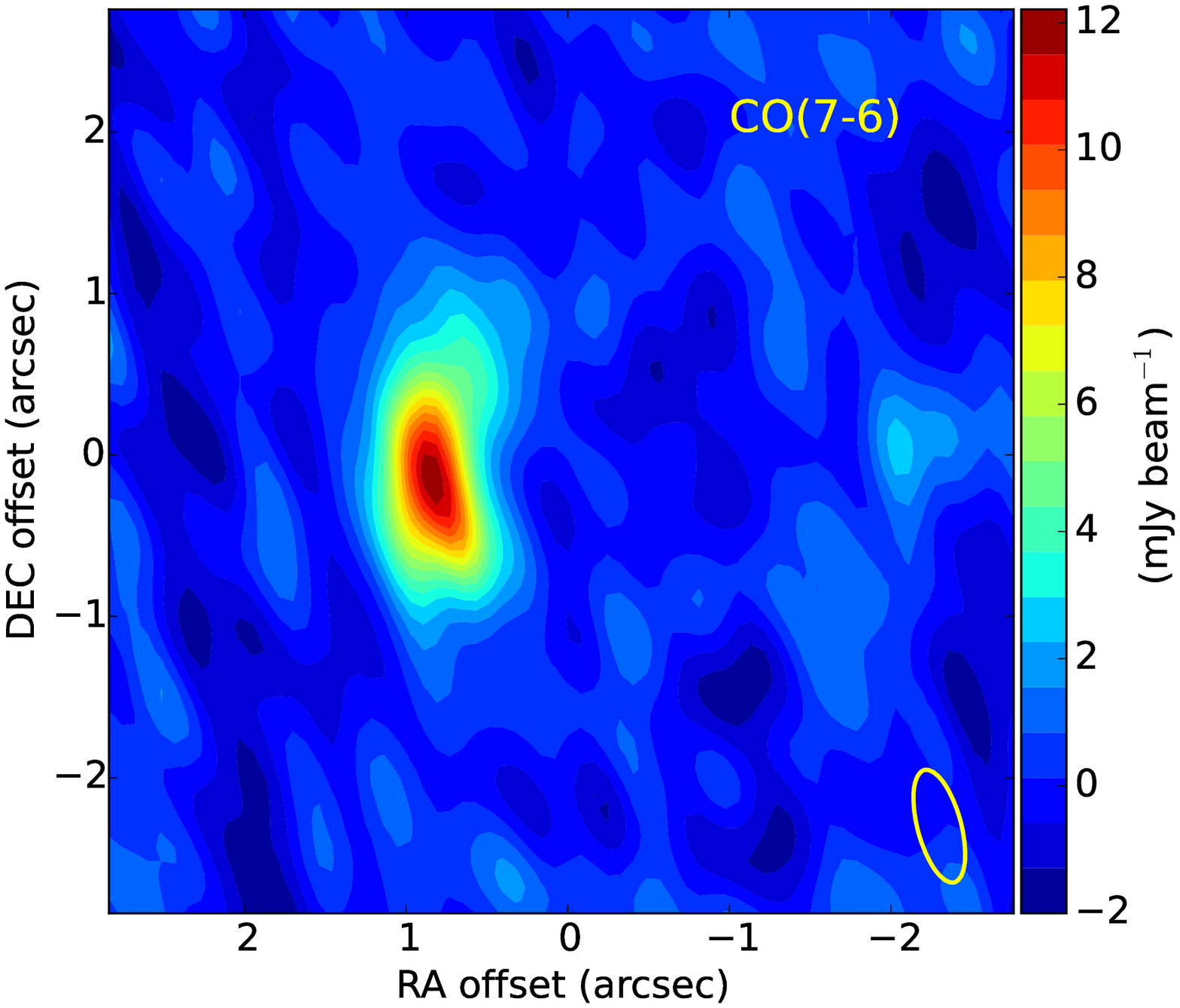}
  \includegraphics[height=4.5cm,trim=0.cm 0.cm 0.cm 0.cm,clip]{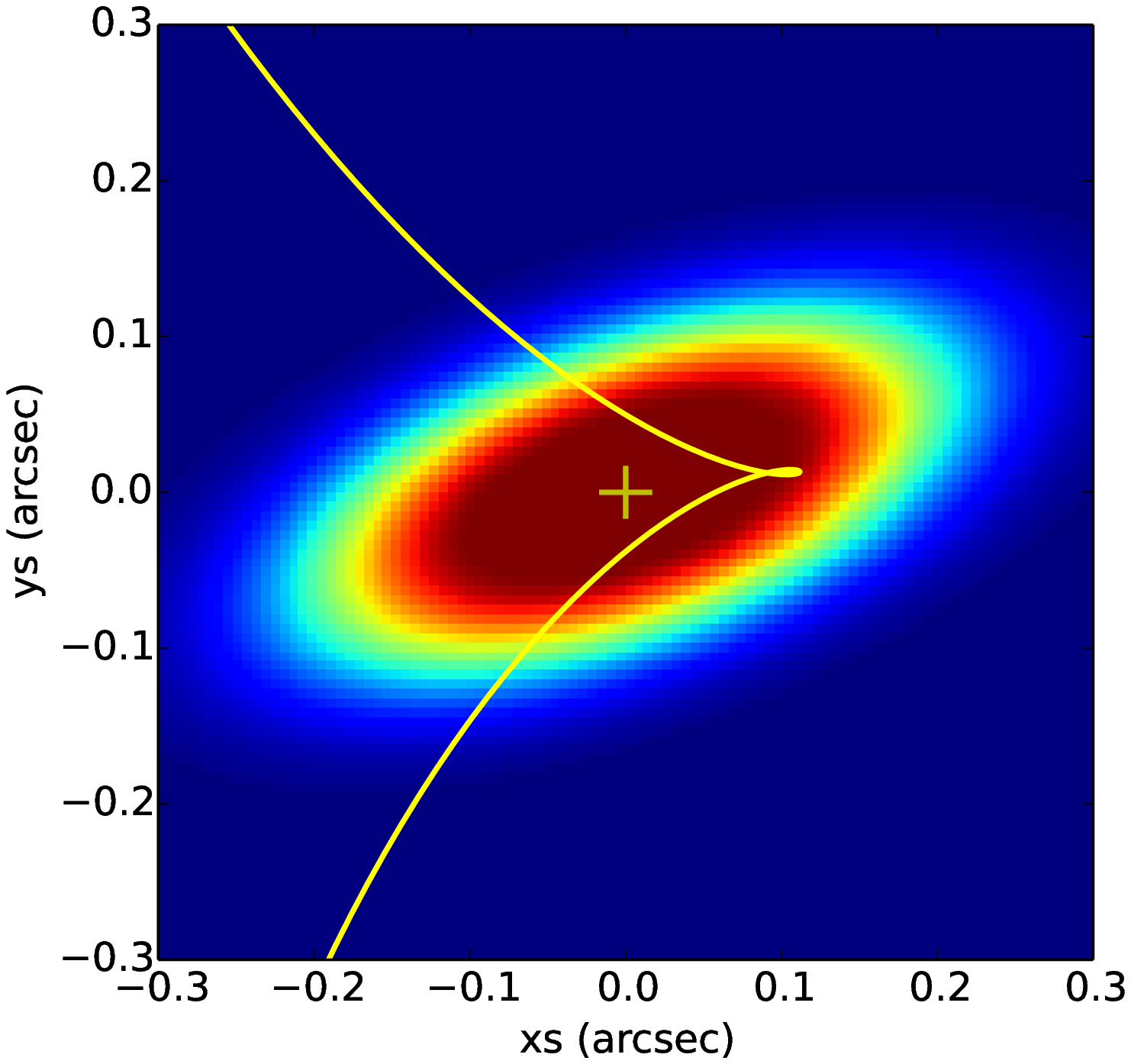}
  \caption{CO(7-6) emission contours of RX J0911 as measured by \citet{Weiss2012} (left) and by \citet{Tuananh2013} (middle, with the lens galaxy at the origin of coordinates). The HST images are superimposed on the left panel. The beams are shown in the lower right corners. Right: brightness distribution in the source plane, as reconstructed by \citet{Tuananh2013}, also showing the caustic and the source position (at the origin of coordinates).}
\label{fig2}
\end{figure*}

Molecular emission has recently been observed in \mbox{CO(1-0)} using the EVLA \citep{Riechers2011d} and in CO(7-6) using the PdBI, first (Figure \ref{fig2} left) by \citet{Weiss2012} and later, with better spatial resolution (Figure \ref{fig2} middle), by \citet{Tuananh2013}. \citet{Hainline2004} mention a probable detection of CO(3-2) emission (to \mbox{3.7 $\sigma$)} but do not resolve the images. As they evaluate a line width a factor more than 3 larger than expected from interpolation between the CO(1-0) and CO(7-6) measurements, we ignore this observation in what follows. An unexpected feature is the narrow width of the line, 120$\pm$14 km\,s$^{-1}$, some four times smaller than for typical quasar hosts at high redshift. The second set of PdBI observations \citep{Tuananh2013,Tuananh2014} made it possible to resolve the gas volume in the source plane as an ellipse (Figure \ref{fig2} right) having an rms radius of \mbox{106$\pm$15 mas} (0.85$\pm$0.12 kpc), an ellipticity parameter (square  root  of  the  ratio  between major  and minor  axes) of $1.60 (^{+0.35}_{-0.18})$ and a position angle (of the major axis) of $111^\circ\pm9^\circ$, 3.3 standard deviations  away  from  a  circular  source  hypothesis. These numbers were obtained from analyses using two different lensing potentials. The differences between their predictions were used to obtain an estimate of the systematic uncertainties. Here we use only one of these, giving the best fit to the HST data, the sum of an elliptic term and a shear term, remembering that it predicts a $\sim$20\% larger rms radius, 129$\pm$15 mas (meaning 1.01$\pm$0.12 kpc or 2.6$\pm$0.3 kpc FWHM), than the average value just quoted. Moreover, evidence  for  a velocity  gradient at the scale of \mbox{$\sim$25 km\,s$^{-1}$\,kpc$^{-1}$} along  the  source  elongation  had  been  obtained  at  the level of 4.5 standard deviations.

The full band X ray luminosity of the quasar has been measured by \citet{Fan2009} as $10^{44.8}$=6.2$\times10^{44}$ erg\,s$^{-1}$; using a relation between dynamical and black hole masses \citep{Bothwell2013} and between X ray luminosity and black hole mass \citep{Alexander2005a,Alexander2005b}, one estimates a black hole mass of $10^{8.2}$=1.6$\times10^8$ solar masses while, on average \citep{McLure2004,Ueda2015}, black hole masses increase from $10^8$ solar masses at $z\sim 0.2$ to $10^9$ solar masses at $z\sim 2$ and between $10^9$ and $10^{10}$ solar masses for $z$>6 \citep{Wang2010}.

Together with the low values \citep{Tuananh2014} of the gas mass (\mbox{$\sim 3.9\times10^9$} solar masses), of the dust mass (\mbox{$\sim 1.3\times10^8$} solar masses) and of the dynamical mass (\mbox{$\sim 4.7\times10^9$} solar masses), the low mass of the central black hole makes RX J0911 an atypical quasar host, a kind of scaled down version of typical quasar hosts at redshifts $\sim$3.

\begin{figure*}
  \centering
  \includegraphics[height=5cm,trim=0.cm 0.cm 0.cm 0.cm,clip]{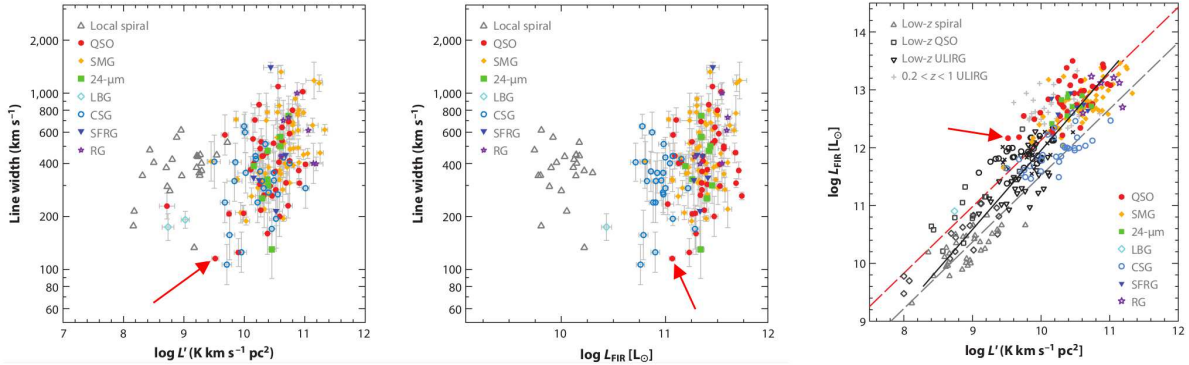}
\caption{CO line width (FWHM) versus CO line luminosity (left), CO line width (FWHM) vs FIR luminosity (middle) and FIR luminosity vs  CO luminosity (right). The red full circles are for quasar hosts. Other galaxy types are specified in the inserts. The red arrow points to RX J0911. From \citet{Carilli2013}.}
\label{fig3}
\end{figure*}

In the present article we complement these earlier studies of RX J0911 with the analysis of recent ALMA observations. In addition to providing new information on the emission from other lines, they contribute, for the first time, high sensitivity data in the 358 GHz continuum.

\section{New observations and data reduction}
\begin{figure}
\centering
\includegraphics[width=6.cm,trim=0.3cm 0.cm 0.cm 0.cm,clip]{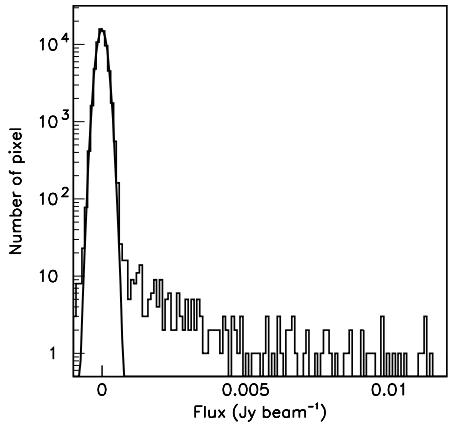}
\caption{Distribution of the flux density (Jy/beam) measured in the continuum. The fit is a Gaussian of mean $-2.6\,\mu$Jy\,beam$^{-1}$ and sigma 165 $\mu$Jy\,beam$^{-1}$.}
\label{fig4}
\end{figure}

We use ALMA observations, number 2011.0.00307.S (PI A. Weiss), performed on 16th November 2012, presently archived and open to public access. The continuum was observed at 358 GHz over four frequency bands for a total on-source time of 26 minutes. It made use of 26 antennas with a maximum baseline of $\sim$ 370 m corresponding to a nearly circular beam of $0.55\times 0.51$ arcsec$^2$. The data have been processed by the ALMA staff and are available in the form of a clean map of 256$\times$256 pixels of 0.1$\times$0.1 arcsec$^2$ each. We checked the good quality of the data reduction, including calibration. We use $x$, $y$ and $z$ coordinates pointing respectively east, north and along the line of sight (away from the observer). The distribution of the mean flux density measured in each pixel is displayed in Figure \ref{fig4} and reveals an effective noise peak having a FWHM of 0.39 mJy\,beam$^{-1}$.  

Also available are lower sensitivity observations of the emission of two CO lines, CO(10-9) and CO(11-10), at respective frequencies of 303.69 and 333.79 GHz. They were observed on 15th November 2011, for on-source times of 10 and 20 minutes respectively, using 16 antennas with a maximum baseline of 126 m (ALMA cycle 0 compact configuration). The beams were nearly circular with \mbox{$\sim$1.5 arcsec} FWHM. 

\section{Clean maps and concentricity of the sources}

\begin{figure}
  \centering
  \includegraphics[height=4.5cm,trim=1.cm 0.cm 1.cm 0.cm,clip]{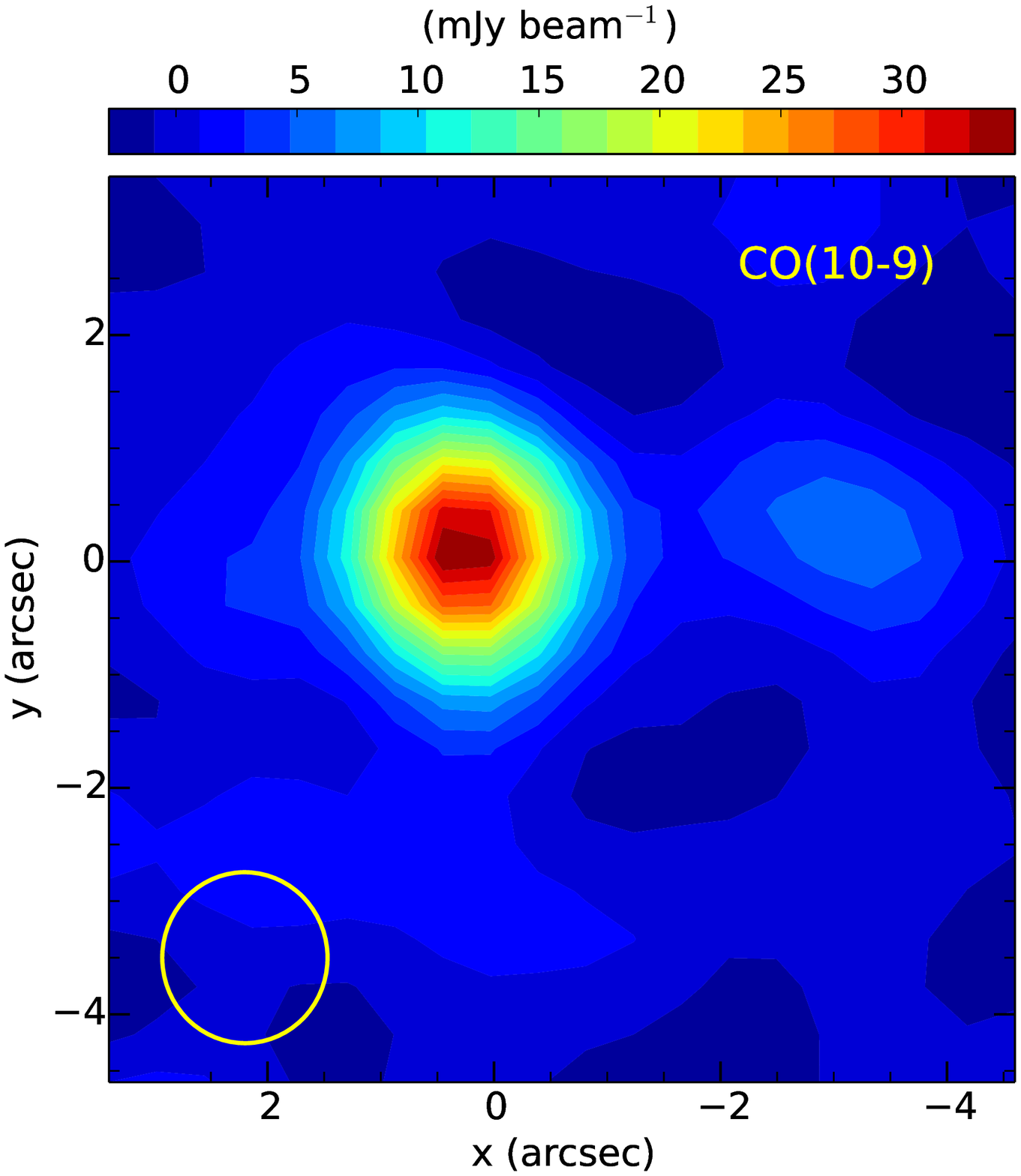}
  \includegraphics[height=4.5cm,trim=1.cm 0.cm 1.cm 0.cm,clip]{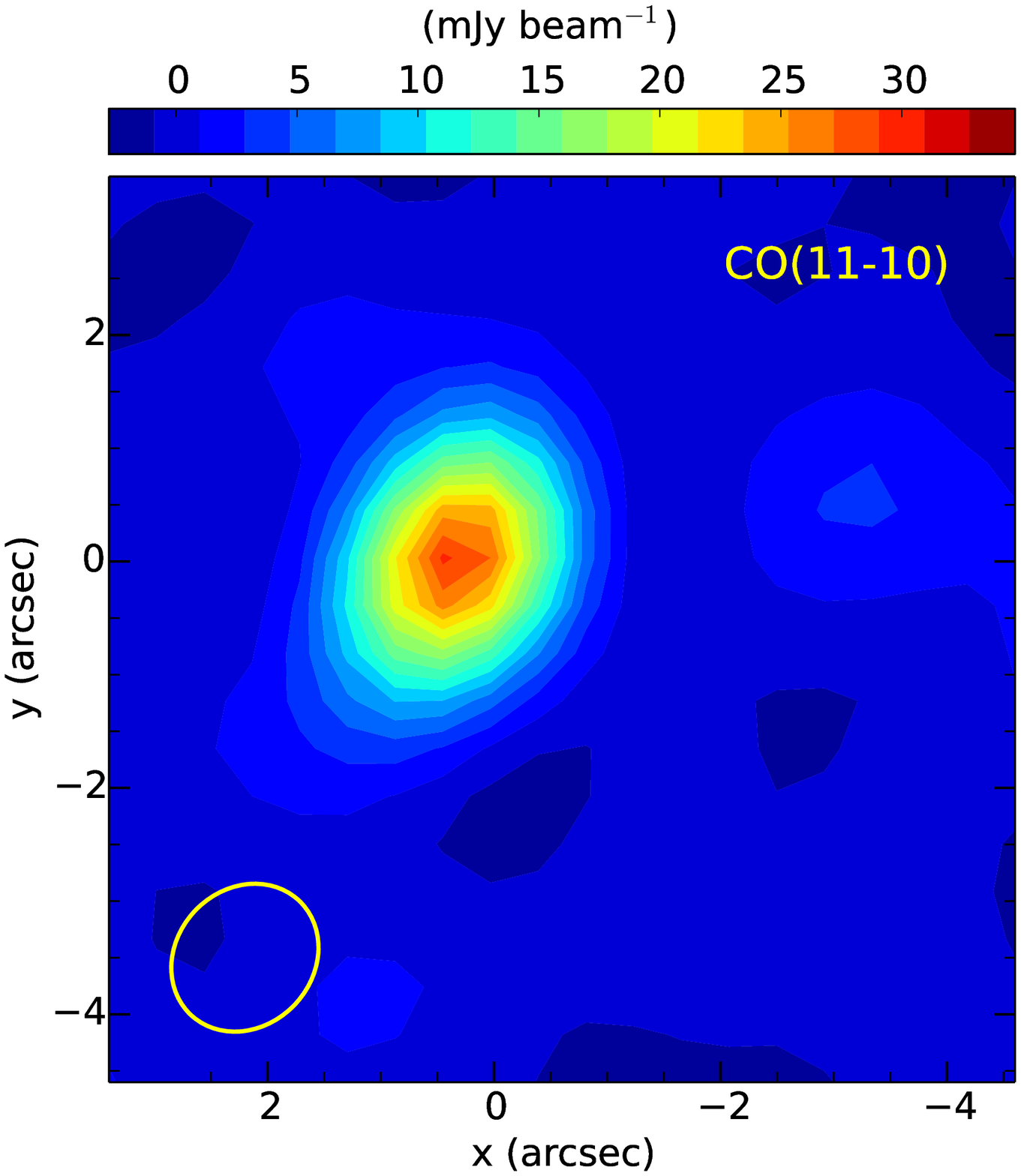}
  \includegraphics[height=5.cm,trim=0.cm 0.cm 0.cm 0.cm,clip]{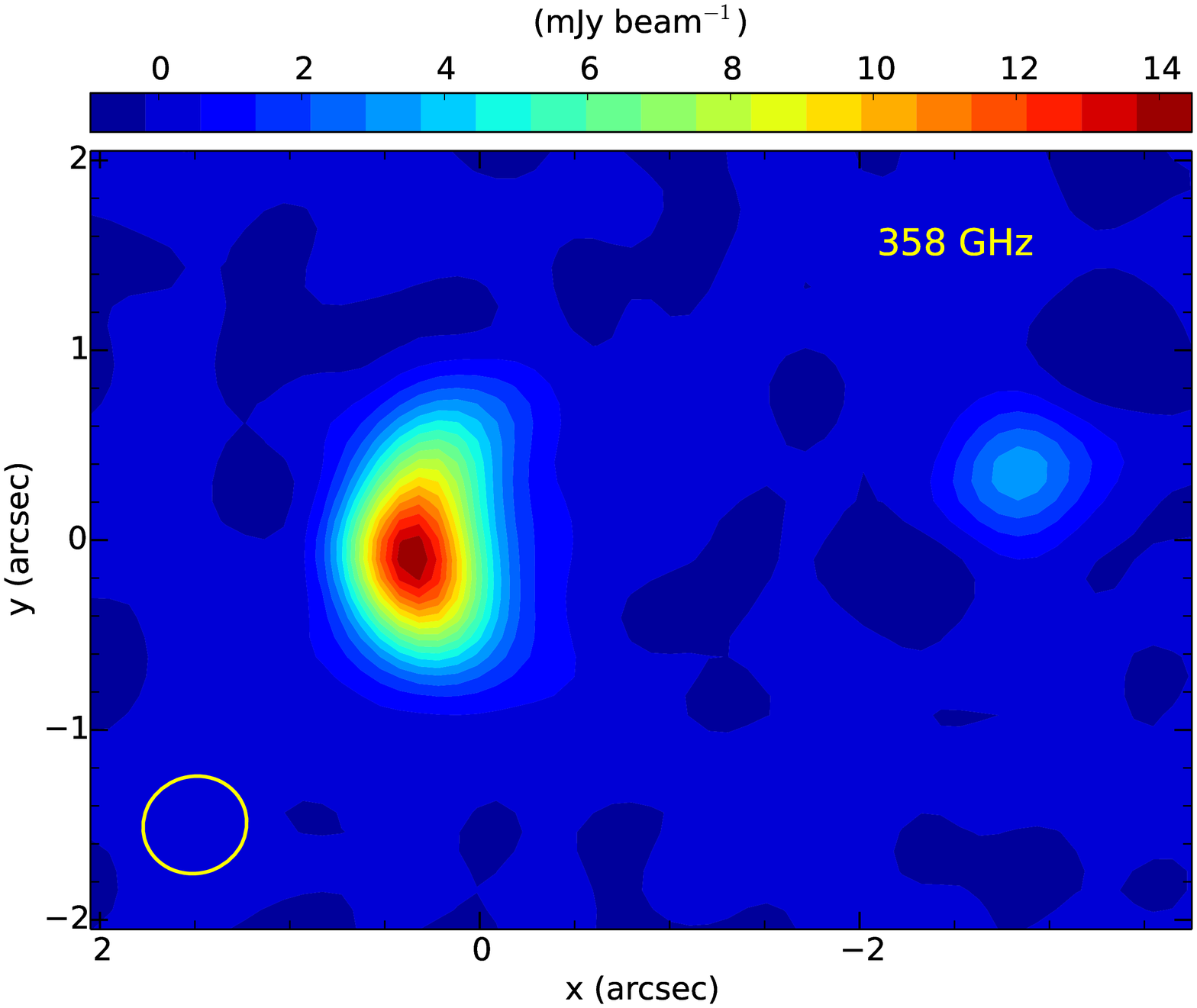}
\caption{Upper panels: clean maps (from left to right) of \mbox{CO(10-9)} and CO(11-10) line emissions. Lower panel: clean map of the 358 GHz continuum. In all panels, the beams are shown in the lower left corners. }
\label{fig5}
\end{figure}

Clean maps of the 358 GHz continuum and of the two molecular line emissions are displayed in Figure \ref{fig5}. For the first time, image B is clearly detected in the continuum, with a signal to noise ratio of $\sim$10 instead of 2.4 in \citet{Tuananh2013}. The new line data have been observed for the first time but the sensitivity and spatial resolution are insufficient to spatially resolve the source.

The gas and dust components may differ in position from the central black hole; moreover, while the latter is essentially a point source, the former are extended sources with different dimensions. It is important, in particular for revealing a possible contribution of mergers, to measure as accurately as possible the positions and dimensions of the sources in each of the three sets of data: HST/optical for the black hole, PdBI/CO(7-6) for the gas component and ALMA/358 GHz for the dust component. An accurate evaluation of possible offsets between the frames of reference associated with each of the three observations is given in Appendix A. While important for ensuring good control over the analysis of the data, it has little influence on the evaluation of possible offsets between the positions of the three sources because of the complexity of the astrometry in a strong lensing configuration such as RX J0911. When the source moves across the lens on the sky plane, there is no simple relation between the global displacement of the images and the displacement of the source. When the source is far away from the lens, the images become faint and fade out; when the source comes close to the lens, it is the configuration of the four (or two) images rather than their global displacement that changes rapidly. This is illustrated in Table \ref{Table2} that lists the displacements $\Delta x_{image}$ and $\Delta y_{image}$ induced on each of the four point images by a displacement ($\Delta x_{source}$,$\Delta y_{source}$) of the point source.

\begin{figure}
\centering
\includegraphics[height=4.cm,trim=1.5cm 1.cm 1.cm 1.cm,clip]{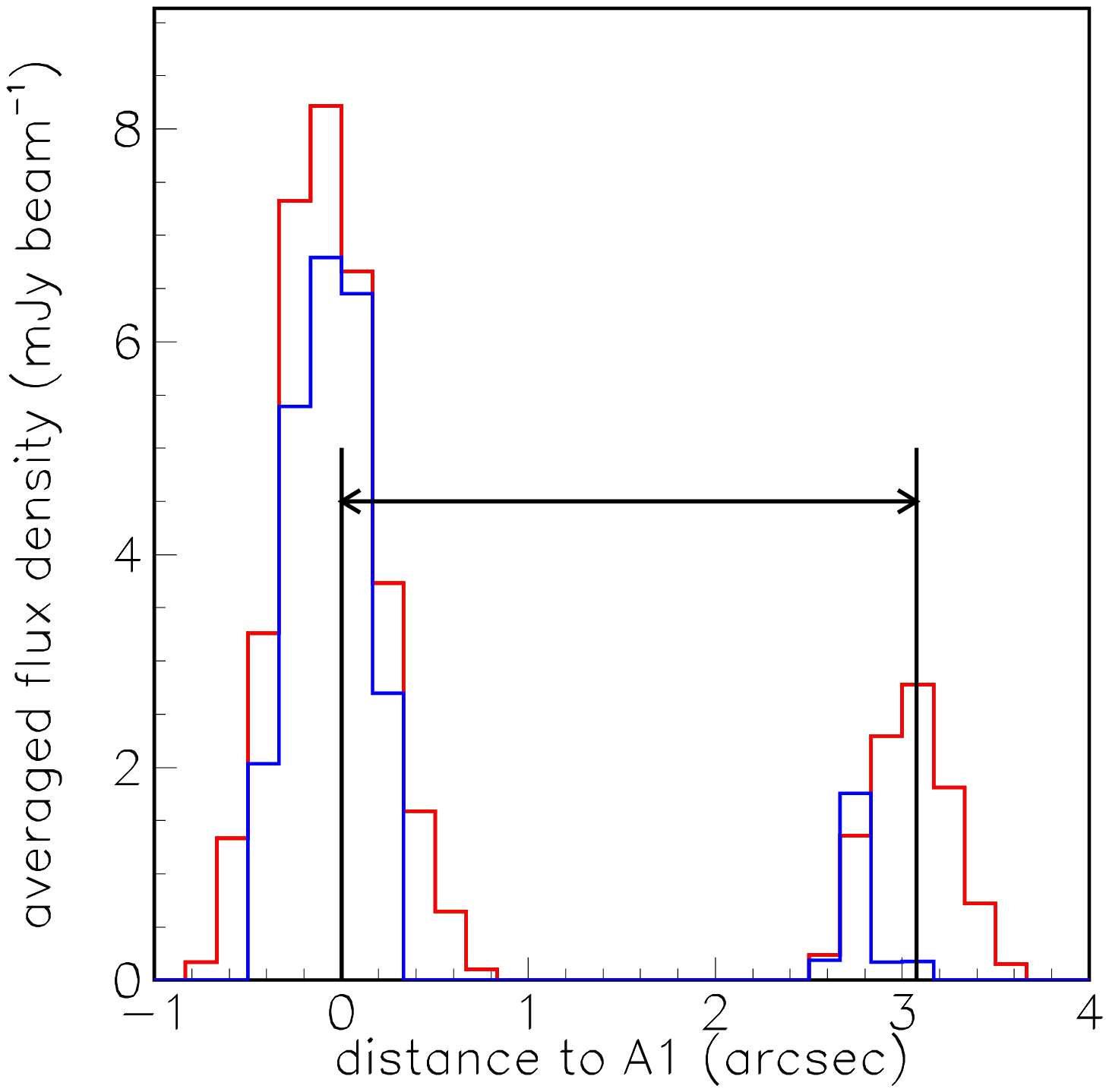}
\includegraphics[height=4.cm,trim=.5cm 1.cm 1.cm 1.cm,clip]{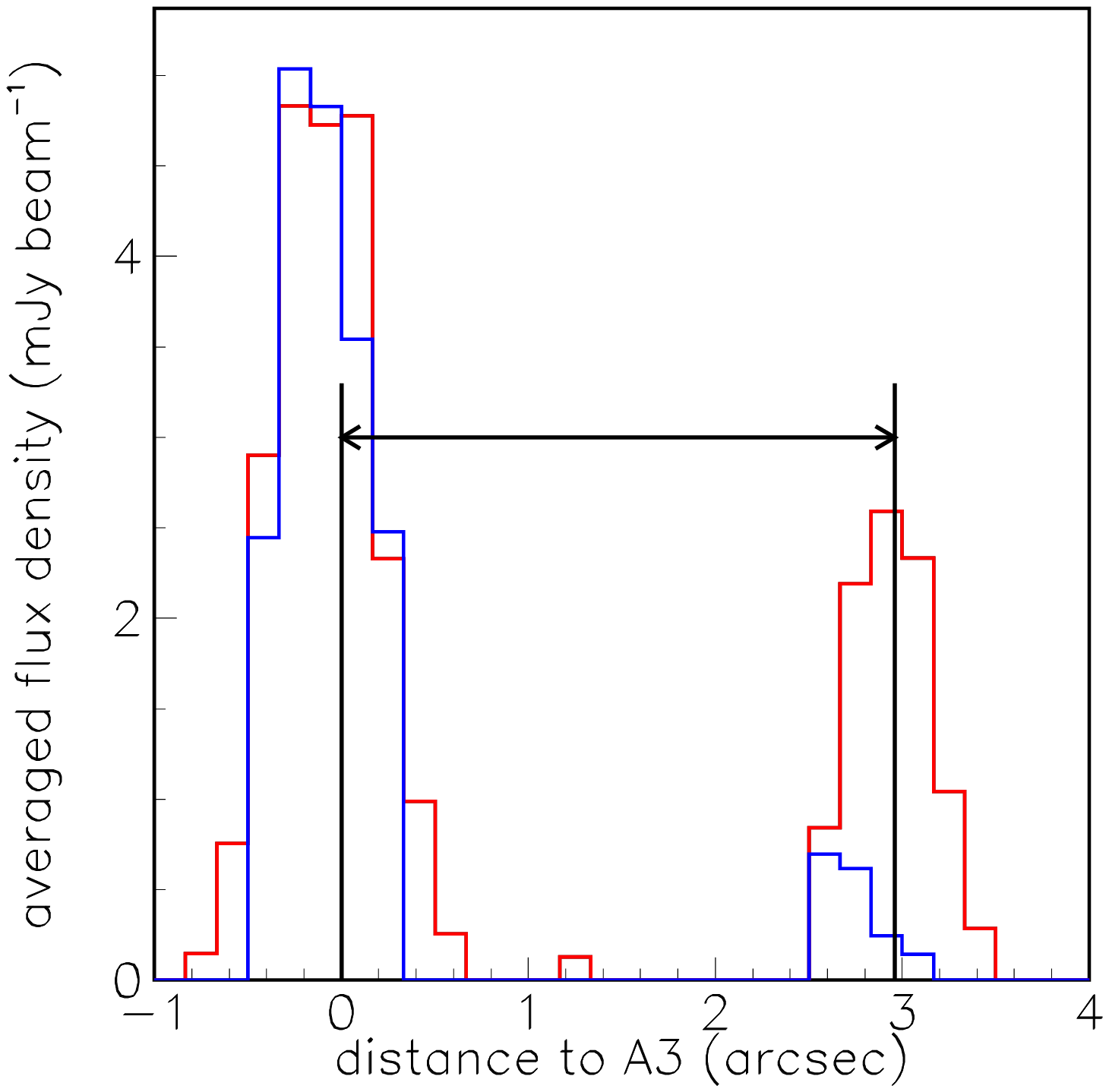}
\caption{Measured distances between image B and A1 (left panel) or A3 (right panel). The blue histograms are for CO(7-6) PdBI data, the red histograms for 358 GHz ALMA data and the black arrows for the quasar HST data.}
\label{fig6}
\end{figure}

\begin{table*}
\centering  
\caption{Displacements induced on each of the four point images by a displacement of the point source (all displacements are measured in milliarcseconds).}   
\label{Table2}        
\begin{tabular}{|c|c|c|c|c|c|c|c|c|c|}
\hline 
$\Delta x_{source}$ & $\Delta y_{source}$ & $\Delta x_{A1}$ & $\Delta y_{A1}$ & $\Delta x_{A2}$ & $\Delta y_{A2}$ & $\Delta x_{A3}$ & $\Delta y_{A3}$ & $\Delta x_{B}$ & $\Delta y_{B}$ \\
\hline 
$-$10 & 0 & 21 & 25 & $-$15 & $-$6 & 11 & $-$23 & $-$15 & 1 \\
\hline
10 & 0 & $-$20 & $-$22 & 15 & 5 & $-$11 & 22 & 14 & 0 \\
\hline
0 & $-$10 & 25 & 8 & $-$8 & $-$55 & $-$21 & 12 & 0 & $-$12 \\
\hline
0 & 10 & $-$22 & $-$6 & 3 & 55 & 24 & $-$15 & $-$1 & 12 \\
\hline
\end{tabular} 
\end{table*}
In particular, we see that the distance between B and A2 varies little when the source moves while the distance between B and A1 increases by 28 mas when the source moves 10 mas west and increases by 19 mas when the source moves 10 mas south. Similarly, the distance between B and A3 increases by 24 mas when the source moves 10 mas west and decreases by 20 mas when the source moves 10 mas south. Writing 
\setlength{\belowdisplayskip}{0pt} \setlength{\belowdisplayshortskip}{0pt}
\setlength{\abovedisplayskip}{0pt} \setlength{\abovedisplayshortskip}{0pt}
\begin{subequations}
  \begin{align}
    \Delta_{BA1}=-2.8 \Delta x_{source}-1.9\Delta y_{source}\\
    \mbox{and}\,     \Delta_{BA3}=-2.4\Delta x_{source}+2.0\Delta y_{source}
    \end{align}
\end{subequations}
we obtain:
\begin{subequations}
   \begin{align}
     \Delta x_{source}=-0.20 \Delta_{BA1}-0.19\Delta_{BA3} \\
     \mbox{and}\,    \Delta y_{source}=-0.24 \Delta_{BA1}+0.28\Delta_{BA3}
   \end{align}
\end{subequations}
We average the 358 GHz continuum flux over two roads along the BA1 and BA3 lines, each $\pm$0.3 arcsec wide, and display in Figure \ref{fig6} the dependence of the results on the distance measured along the road, giving \mbox{$\Delta_{BA1}=37$ mas} and $\Delta _{BA3}=55$ mas from which we obtain \mbox{$\Delta x_{source}=-18 \pm 14$ mas} and $\Delta y_{source}=7 \pm 18$ mas. The errors include uncertainties on both the distance measurements and the parameters of the pair of linear equations.

Repeating the exercise on the CO(7-6) PdBI data (Figure \ref{fig6}) gives \mbox{$\Delta_{BA1}=-268$ mas} and \mbox{$\Delta_{BA3}=-118$ mas} from which we obtain \mbox{$\Delta x_{source}=76 \pm 41$ mas} and \mbox{$\Delta y_{source}=31 \pm 53$ mas}. In this case, the larger uncertainties are due to the low signal to noise ratio of image B.

We obtain upper limits on the distance of the source from the origin by adding the $x$ and $y$ contributions in quadrature, each individual contribution being the sum of the value and its uncertainty. As a result, we retain from this analysis that the quasar and 358 GHz emissions are concentric within 40 mas, meaning 0.31 kpc, while the quasar and CO(7-6) emissions are concentric within 140 mas, meaning 1.10 kpc.

These results do not depend on the offsets between the three frames of reference, which are evaluated in Appendix A. The result of this latter evaluation is expressed in terms of the origins of the PdBI and HST frames of coordinates as measured in the ALMA frame:\\
$-$ for the PdBI: $-25\pm 40$ mas in $x$ and $-117\pm 80$ mas in $y$;\\
$-$ for the HST: 122 mas in $x$ and 138 mas in $y$ with uncertainties poorly defined but at the 100 to 200 mas scale.

\section{Measurement of the source sizes}
\begin{figure}
\centering
\includegraphics[width=.22\textwidth,trim=1.5cm 1.cm 1.cm 1.cm,clip]{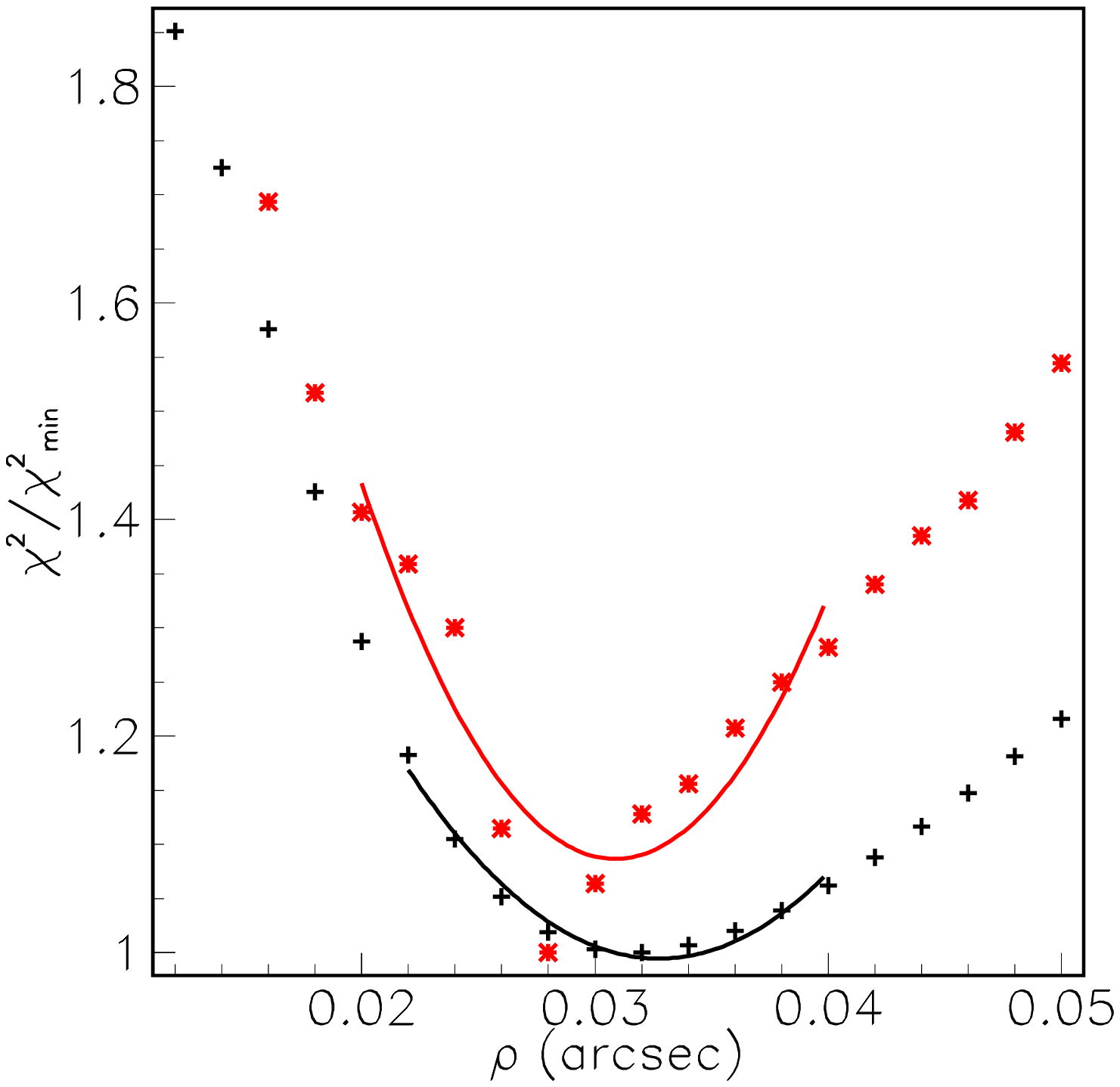}
\includegraphics[width=.25\textwidth,trim=0.cm 0.1cm 0.cm 0.cm,clip]{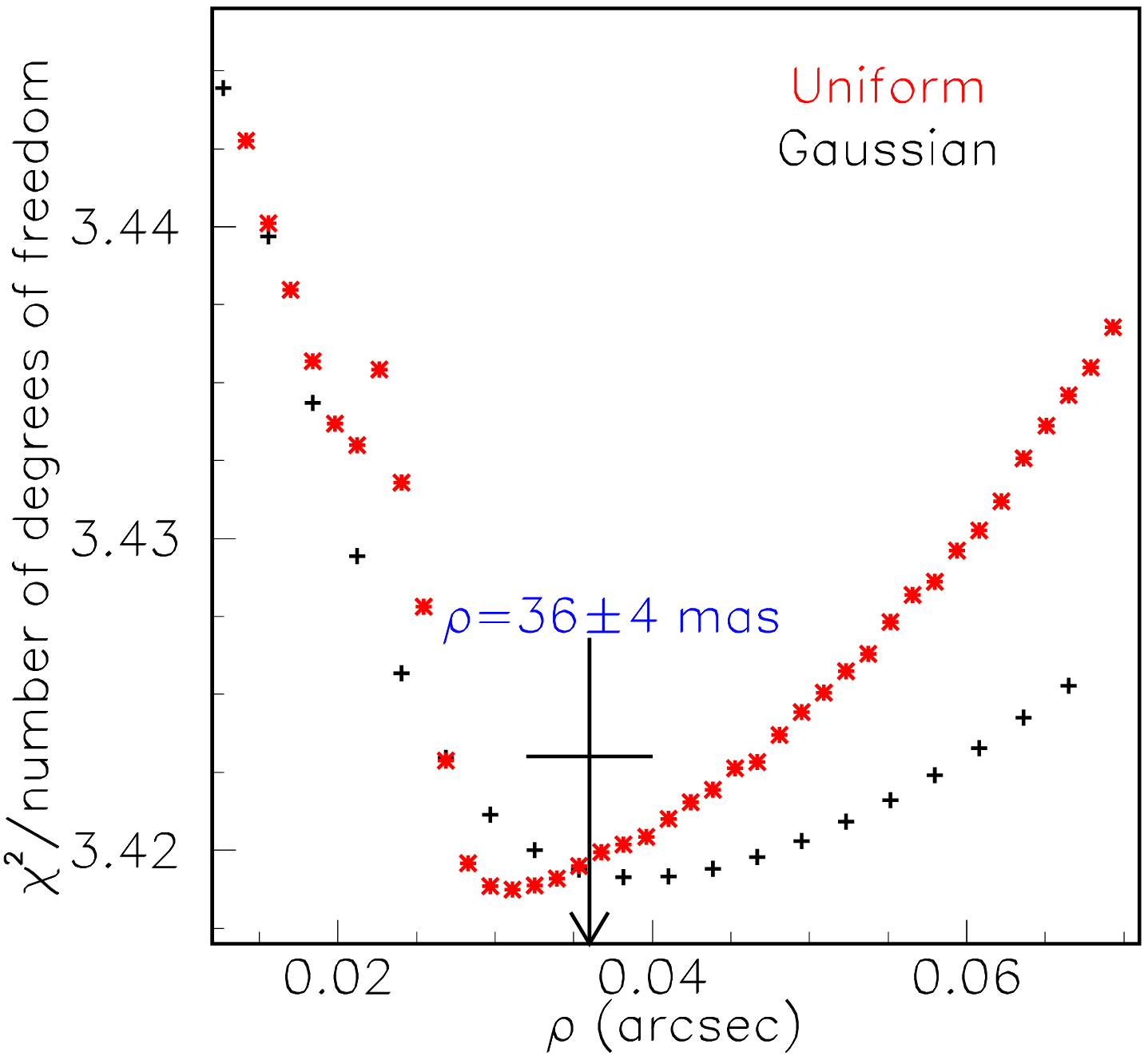}
\caption{Left: dependence on the source size (arcsec) of the  $\chi^2$ (normalised to minimum) of the fit to the data on the clean map. Both the uniform disc (red) and Gaussian profile (black) cases are illustrated. The source is positioned at its nominal position (the values quoted in Table \ref{Table3} are obtained by averaging over source positions within $\pm$10 arcsec). The curves are parabola fits. Right: dependence on the source size (arcsec) of the  $\chi^2$ (normalised to the number of degrees of freedom) of the fit to the data in the $uv$ plane. Both the uniform disc (red) and Gaussian profile (black) cases are illustrated. The blue arrow shows the value \mbox{retained for $\rho$.}}    
\label{fig7}
\end{figure}

In order to obtain the morphology of the continuum source and to measure its size, we use the lensing potential described in Table \ref{TableA1}, correct for the offsets listed in Table \ref{TableA2} and assume a circular morphology in the form of either a uniform disc or a Gaussian, two extreme configurations that may reasonably be thought to bracket reality. The fits are first made on the clean map, using as uncertainty 10\% of the measured flux density added in quadrature to the 165 $\mu$Jy\,beam$^{-1}$ of effective noise (Figure \ref{fig4}). This procedure is not rigorous because the measurement uncertainties affecting neighbour pixels are strongly correlated. For this reason, a final fit is ultimately made in the $uv$ plane in order to overcome this difficulty. Table \ref{Table3} summarizes the result of the best fits and Figure \ref{fig7} displays the dependence of their $\chi^2$ on the source size defined as the root mean square radius measured over the source area, $\rho=\sqrt{<R^2>}=\sqrt{\frac{\int R^2F(x,y)\,dx\,dy}{\int F(x,y)\,dx\,dy}}$ where we recall that $x$ points east, $y$ points north, $z$ points along the line of sight away from the observer, $F(x,y)$ is the integrated flux density measured in pixel ($x$,$y$) and $R=\sqrt{x^2+y^2}$. The values taken by the ratio of $\rho$ to the half width at half maximum are $\frac{\sqrt{2}}{2}\sim 0.71$ for a uniform disc and $\sim$0.85 for a Gaussian profile. A same value of 34$\pm$6 mas (0.27$\pm$0.05 kpc) is obtained for $\rho$ in the continuum ALMA data for both the uniform disc and Gaussian profile configurations; it corresponds therefore to a FWHM of 0.68$\pm$0.12 kpc, in agreement with the much less precise estimate of \citet{Tuananh2014}. The uncertainty attached to this measurement has been evaluated for an uncertainty of $\pm$10 mas of the source position in both $x$ and $y$ \citep{Tuananh2014}.

\begin{table*}
\centering  
\caption{Summary of source size measurements. Estimated uncertainties (1 s.d.) are indicated in parentheses.}
\label{Table3}        
\begin{tabular}{|c|c|c|c|}
\hline 
Observations & Method & $\rho=\sqrt{<R^2>}$(mas) & FWHM (kpc) \\
\hline 
\multirow{5}{*}{Continuum Present ALMA data}
& Fit uniform disc & \makecell{Clean map: 34(4) \\$uv$ plane: 32} & \makecell{Clean map: 0.68(0.08) \\$uv$ plane: 0.64}\\
\cline{2-4}
& Fit Gaussian profile & \makecell{Clean map: 34(9)\\ $uv$ plane: 40} & \makecell{Clean map: 0.68(0.18) \\$uv$ plane: 0.80} \\
\cline{2-4}
& Fit retained & 36(4)& 0.72(0.08)\\
\cline{2-4}
& B/A & 48(8) & 0.96(0.16)\\
\cline{2-4}
& {\bf{Final retained}} & {\bf{38(4)}} & {\bf{0.76(0.08)}} \\
\hline
\multirow{2}{*}{Continuum \citep{Tuananh2014}}
& disc & 37(14) & 0.7(0.3) \\
\cline{2-4}
& Gaussian & 55(21) & 1.1(0.4) \\
\hline
\multirow{3}{*}{CO(7-6) \citep{Tuananh2014}}
& disc &  115(13) & 2.3(0.3) \\
\cline{2-4}
& Gaussian & 148(20) & 3.0(0.4) \\
\cline{2-4}
& {\bf{ellipse}} & {\bf{129(15)}} & {\bf{2.6(0.3)}}\\
\hline
\end{tabular} 
\end{table*}

As a check of the validity of these results and for a more rigorous treatment of uncertainties, we now repeat the fit in the $uv$ plane rather than on the clean map. The fit is done over 76$\times$10$^3$ complex visibilities, meaning $152\times10^3$ degrees of freedom. The best fit $\chi^2$ is $\sim$3.4 per degree of freedom, suggesting that we underestimate the uncertainty by a factor $\sim$1.8, either because the noise itself is higher than estimated or because another source of error has been neglected. In any case, the uncertainty on the source size is completely dominated by systematic effects, the difference between the uniform disc and Gaussian profile fits providing a good illustration, the respective $\rho$ values being 32 and 40 mas. This result confirms the results of the analyses performed on the clean map and is more reliable because of its better treatment of the uncertainties. We therefore retain as a final result $\rho=36\pm 4$ mas, meaning a FWHM of 0.72$\pm$0.08 kpc, the quoted uncertainty accounting for the sources of systematic errors associated with the source profile and position as well as with the knowledge of the lens.

\begin{figure*}
  \centering
  \includegraphics[height=5 cm,trim=1.cm 0.cm 2.cm 0.cm,clip]{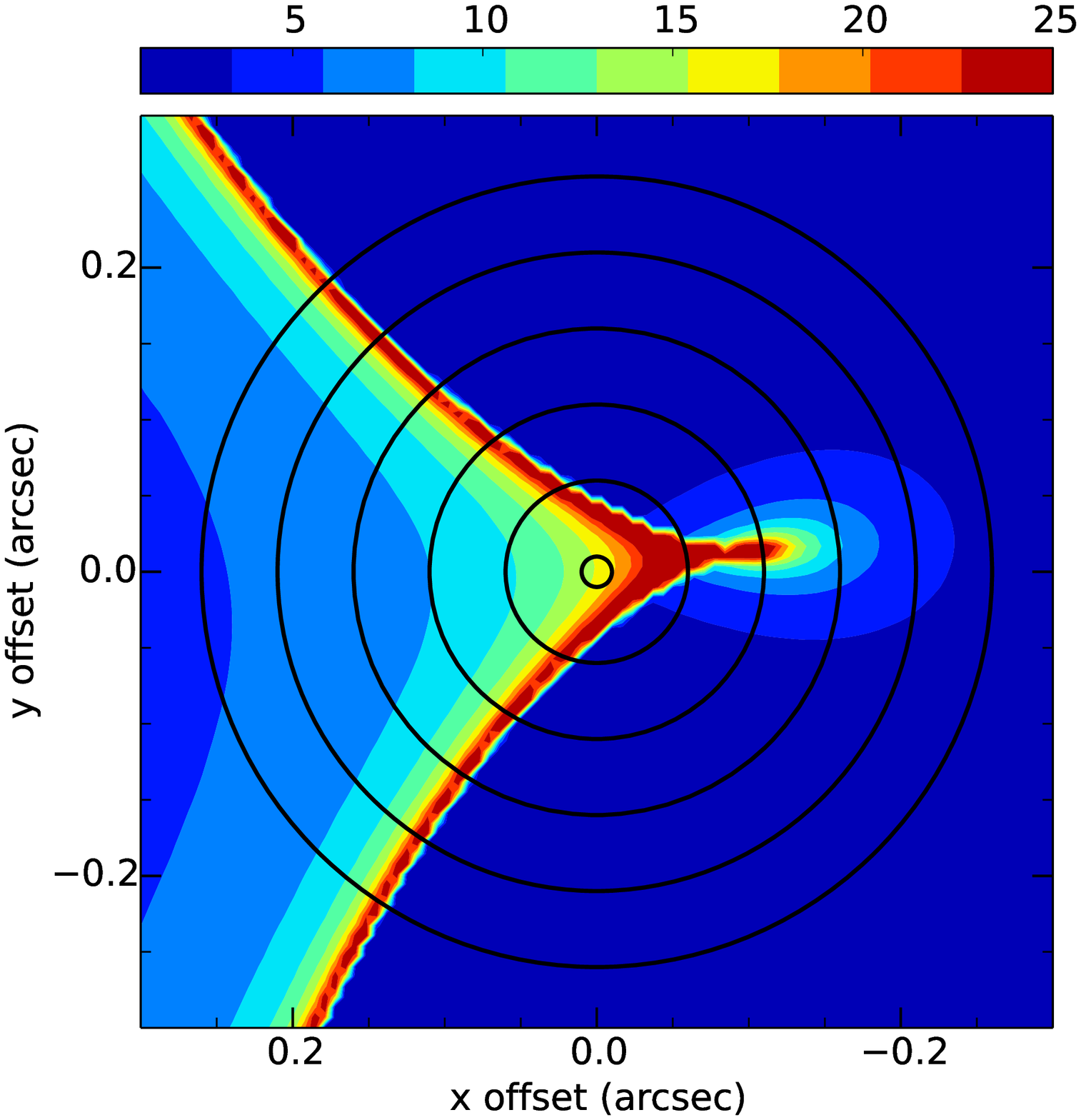}
  \includegraphics[height=5 cm,trim=0.5cm 0.cm 2.cm 0.cm,clip]{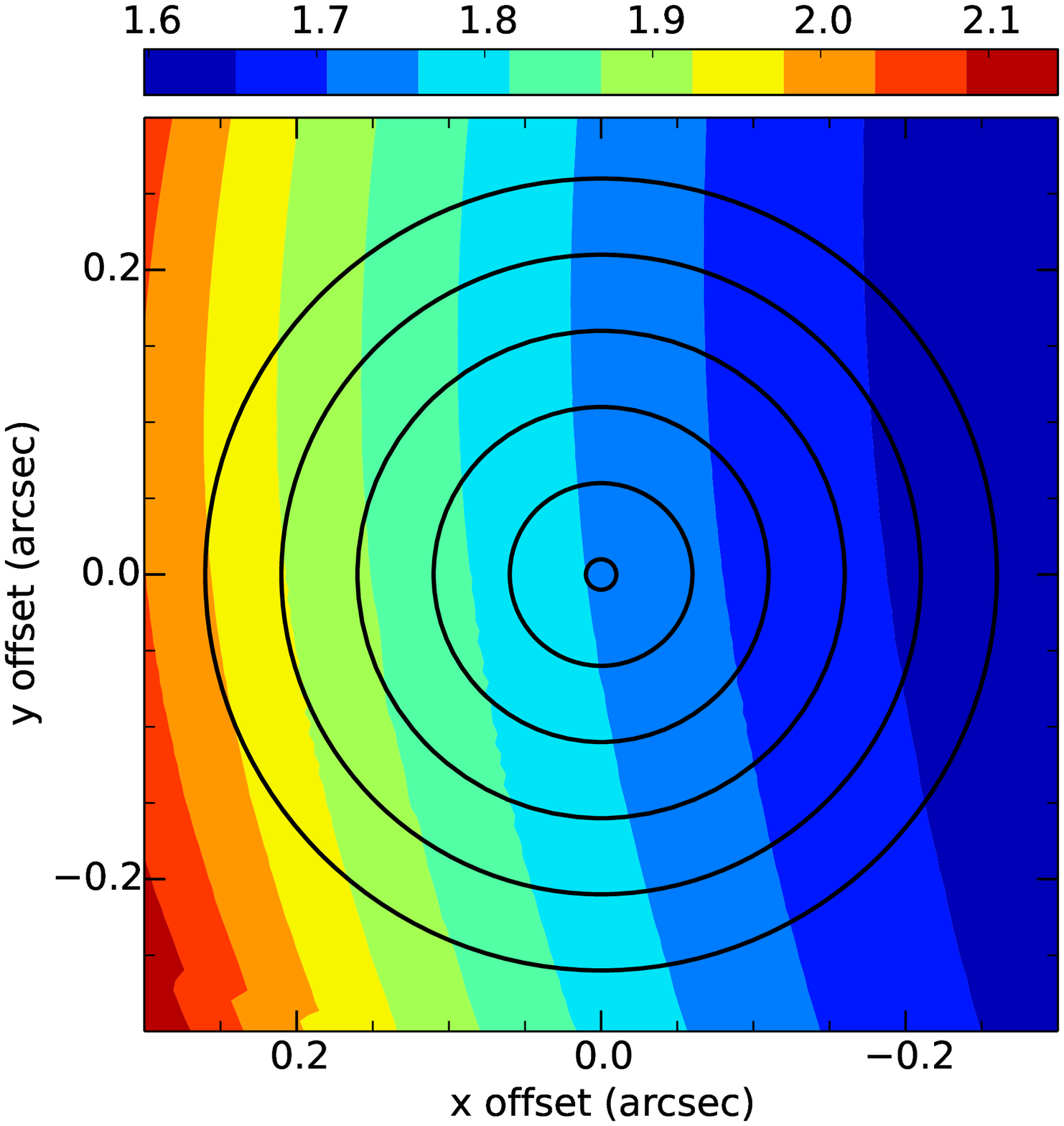}
  \includegraphics[height=4.85 cm,trim=-1.5cm 0.7cm 0.cm 0.cm,clip]{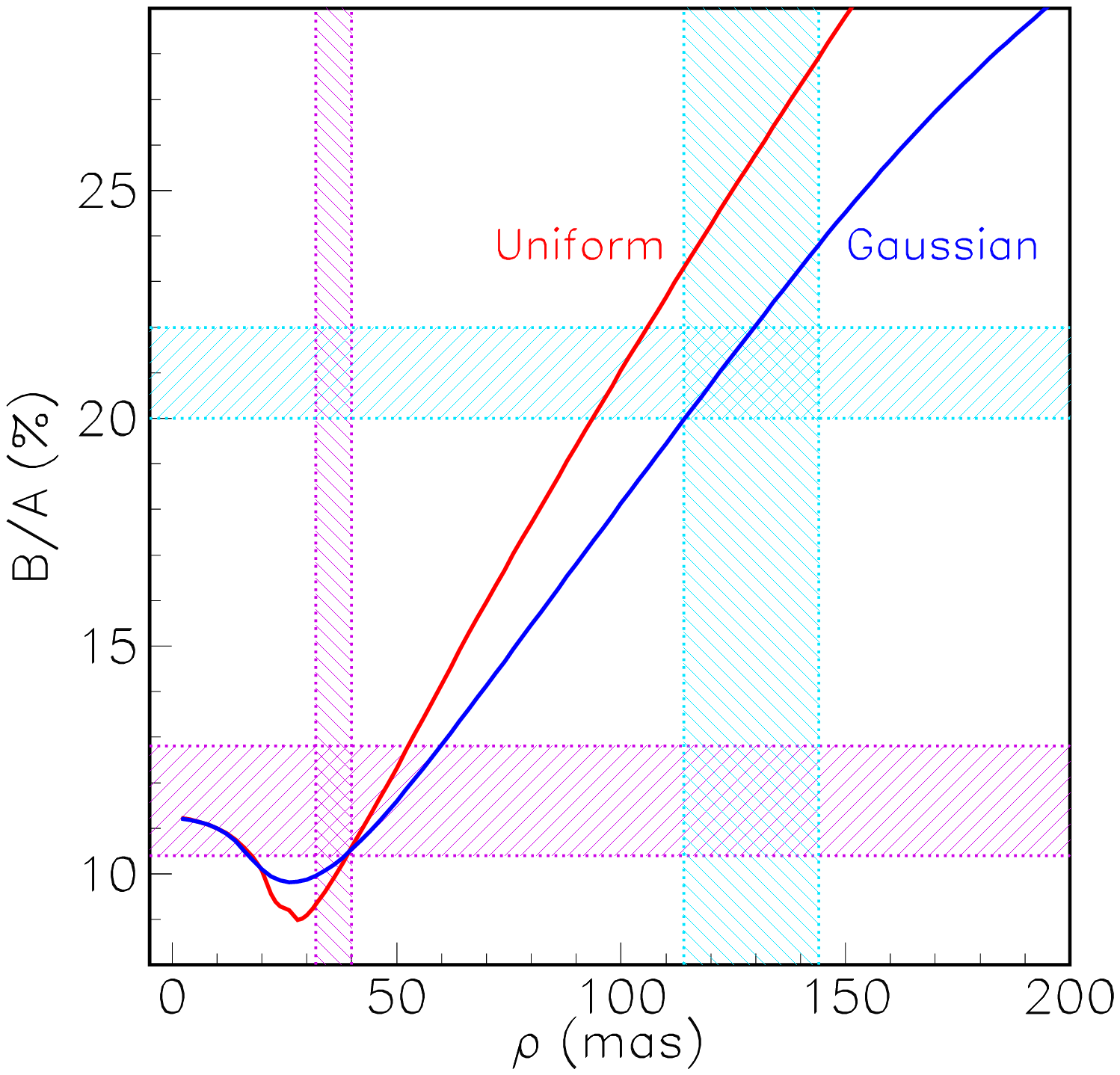}
\caption{Distribution of the magnifications in the source plane for the three A images considered globally (left panel) and for image B (middle panel). Concentric circles show sources of radii increasing from 10 mas in steps of 50 mas. The right panel displays the dependence of the B/A ratio on $\rho$ (mas) for disc sources having either uniform (red) or Gaussian (blue) brightness lensed by the potential described in Table \ref{TableA1}. The bands show the results of measurements ($\pm$1$\sigma$) for both the 358 GHz continuum (purple, this work, $\rho=36\pm 4$ mas; B/A=11.6$\pm$1.2\%) and the CO(7-6) emission (cyan, \citealt{Weiss2012} for the B/A ratio, 21$\pm$1\%; \citealt{Tuananh2014} for the fit in the $uv$ plane, $\rho=129\pm 15$ mas).}

\label{fig8}
\end{figure*}

As explained in \citet{Tuananh2014}, an independent evaluation of the source size can be obtained from the ratio between the observed brightness of images B and A's (considered here globally, namely the flux being integrated over the area covered by the three A images). Point sources located outside the caustic produce only two images, B and A2, while inside the caustic they produce also the A1-A3 pair, adding up to four images. It is clear from Figure \ref{fig1} (centre and right) that the relative occurrence of two-image configurations increases with the size of the source, causing the A images to become globally fainter with respect to image B, the latter being always present in the two-image configuration: the B/A brightness is a measure of the source size. This ratio is essentially unaffected by beam convolution and its measurement does not require a very good angular resolution (yet sufficient to separate A from B). However, as image B is not much magnified, it requires a high sensitivity to be obtained with good precision. This is illustrated in Figure \ref{fig8} (left and centre), which displays separately the variations over the source plane of the A and B magnifications predicted by the model. The A images are faint in most of the two-image region, large magnifications being only reached in the vicinity of the caustic cusp. On the contrary, the magnification of image B, which is far from the critical curve, is always weak, whether the source point is inside or outside the caustic cusp: it varies very slowly across the region explored in Figure \ref{fig8}.
In the HST case, as noted by \citet{Burud1998a,Burud1998b} and their successors, images A suffer an important reddening while image B is unaffected. As a result, the ratio B/A, expected to be $\sim$10\% for a point source, is measured between 18\% and 20\%, preventing the use of the method in this part of the frequency spectrum. However, such an effect should be negligible in the millimetre wavelength region. \citet{Weiss2012} quote a value of 21$\pm$1\% for B/A for CO(7-6) emission but we note that this is the ratio between the maxima of the B and A flux distributions displayed in their Figure \ref{fig2} (left), which one would expect to differ from the ratio between the integrated fluxes. The present 358 GHz ALMA data give a B/A ratio of 11.6$\pm$1.2\%. Figure \ref{fig8} (right) compares these results with the prediction of the lensing model using circular sources having both uniform and Gaussian brightness distributions. The value of $\rho$ obtained in this manner for the 358 GHz continuum is 48$\pm$8 mas, in agreement with the values obtained from the clean map and $uv$ plane fits, respectively 34$\pm$6 mas and 36$\pm$4 mas. Such agreement between results obtained by two completely independent methods brings additional confidence in their validity. Taking all the above in due account, we retain as final value of the size of the 358 GHz continuum emission $\rho=38\pm 4$ mas, meaning 0.30$\pm$0.03 kpc or 0.76$\pm$0.08 kpc FWHM. It is 3.4$\pm$0.4 times smaller than the size of the CO(7-6) emission measured by \citet{Tuananh2014}.

\begin{figure}
\centering
\includegraphics[height=5cm,trim=0.cm 0.cm 0.cm 0.cm,clip]{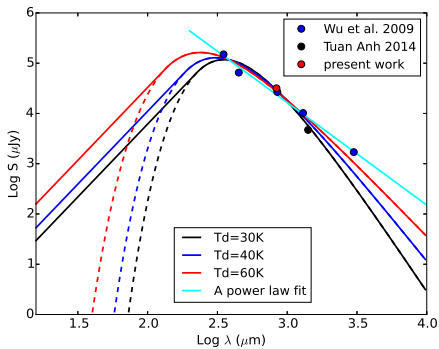}
\caption{Wavelength dependence of the measured dust luminosity of RX J0911 ($\mu$Jy and $\mu$m, decimal exponents on both axes). The present measurement (red) is shown together with earlier data (blue) from \citet{Wu2009}, \citet{Barvainis2002} and \citet{Tuananh2014}. The cyan line is the result of a power law fit corresponding to an emissivity index of 2.04. The curves are the results of black body fits for respective temperatures of 30 K (black), 40 K (blue) and 60 K (red).}
\label{fig9}
\end{figure}

\begin{figure}
\centering
\includegraphics[height=4.cm,trim=0.cm 0.cm 0.cm 0.cm,clip]{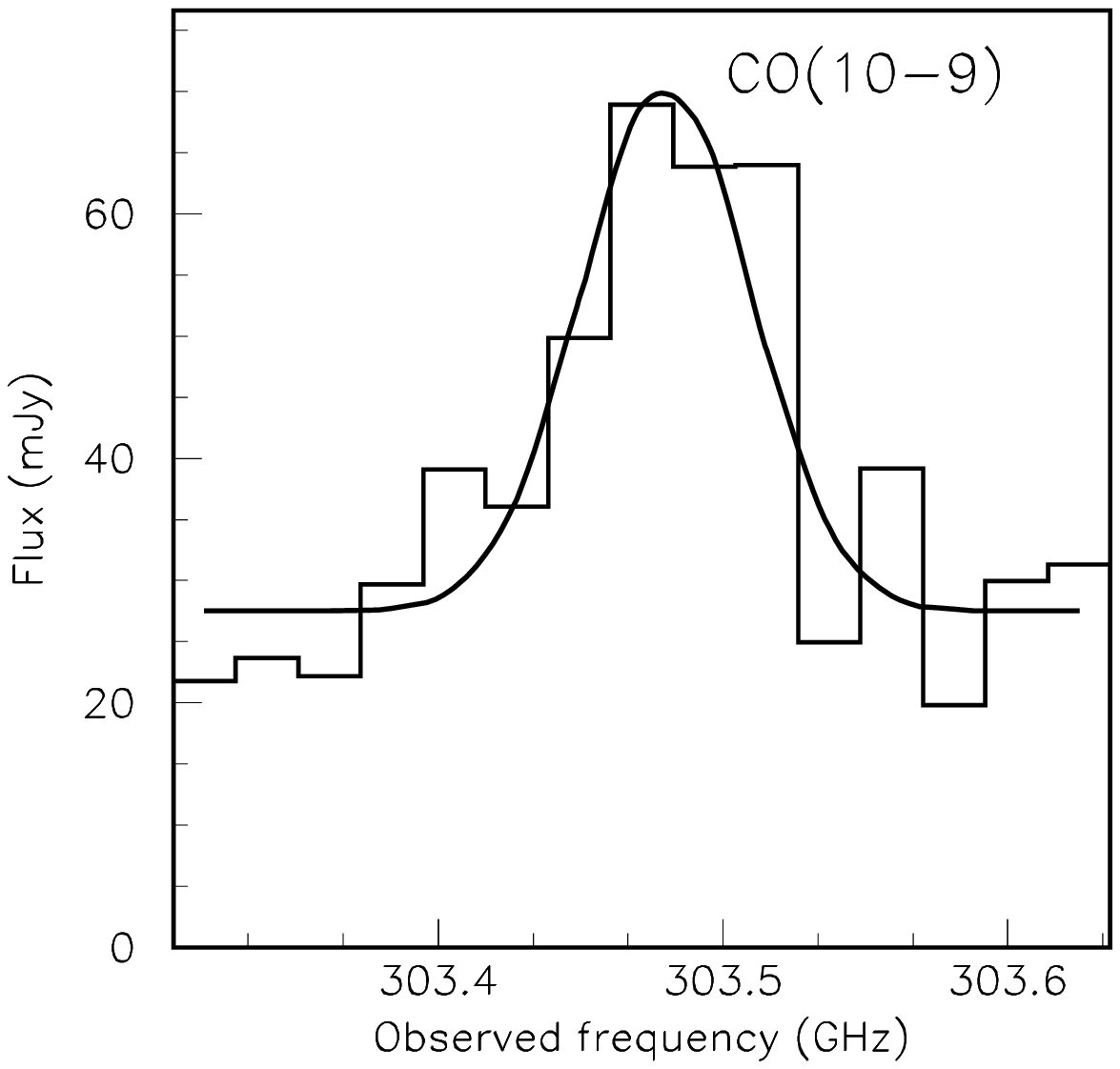}
\includegraphics[height=4.cm,trim=0.cm 0.cm 0.cm 0.cm,clip]{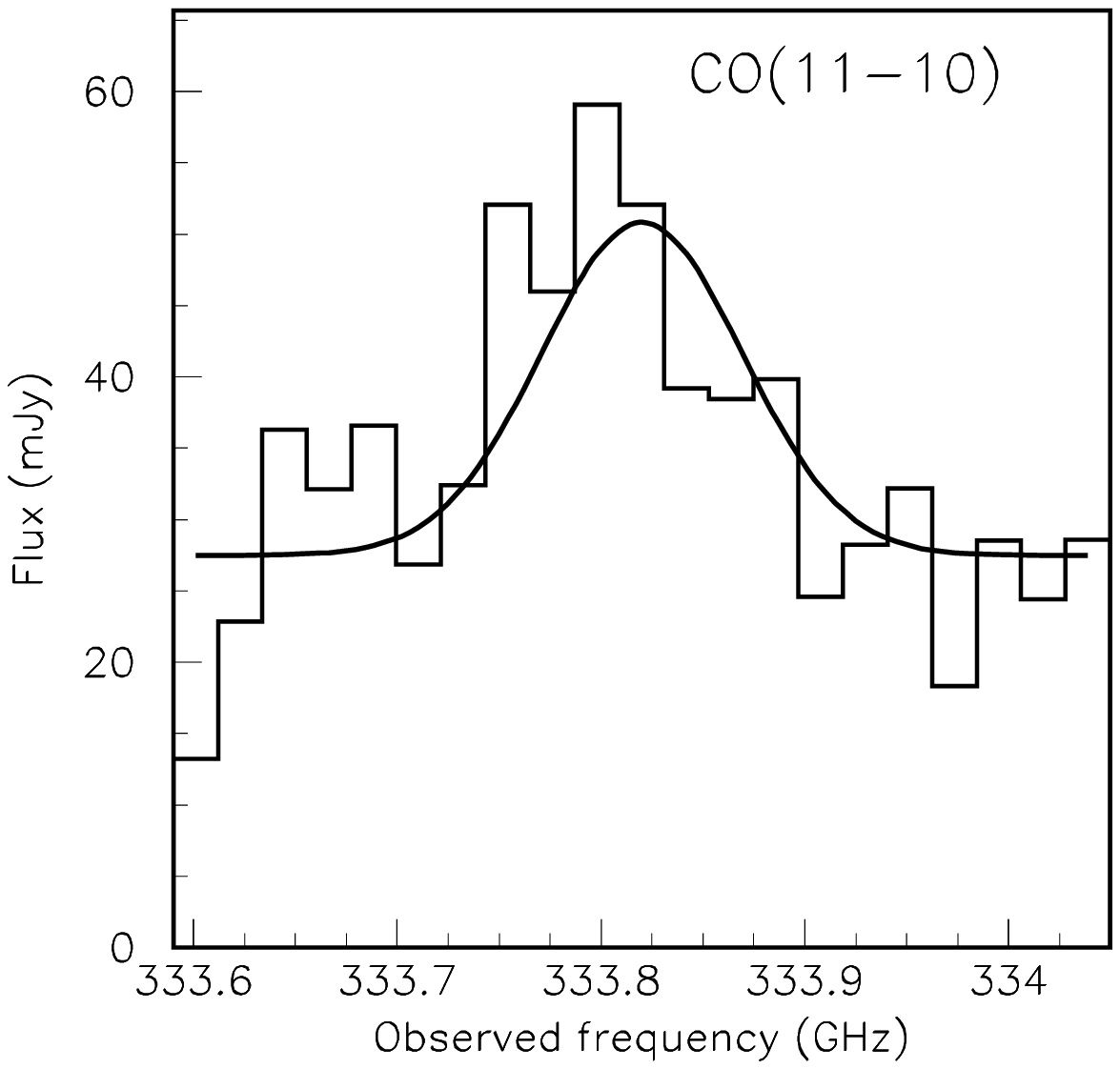}
\caption{Observed line profiles integrated over the region of the A and B images for CO(10-9) (left) and CO(11-10) (right). The lines are Gaussian fits over a continuum level scaled down from the observed 358 GHz measurement. Fluxes (ordinate) are in mJy and frequencies (abscissa) in GHz.}
\label{fig10}
\end{figure}

\section{Integrated continuum flux}
The dust luminosity, evaluated over the regions covered by the A and B images, with an area of 9.4 beams, adds up to 31.7 mJy at an average wavelength of 833 $\mu$m. The uncertainty attached to this number amounts to 0.5 mJy. Figure \ref{fig9} compares this result, 31.7$\pm$0.5 mJy, with earlier measurements, with which it is in excellent agreement. However, it does not help with the evaluation of the dust temperature, being in a frequency range where the black body luminosity is essentially temperature independent. Indeed, the suggestion of a high dust temperature occasionally mentioned in the literature rests on a single measurement of 1.7$\pm$0.3 mJy at 3 mm wavelength by \citet{Barvainis2002}.

\section{Line parameters}
Figure \ref{fig10} displays the observed line profiles of CO(10-9) and CO(11-10). We use the evaluation of the continuum displayed in Figure \ref{fig9} to evaluate by scaling the continuum contributions as 22.7 mJy and 27.5 mJy respectively. Subtracting such contributions and fitting Gaussian profiles gives the results listed in Table \ref{Table4}. The extreme narrowness of the line previously observed in CO(1-0) and CO(7-6) emission is confirmed. The observed frequency values are in excellent agreement with expectation within errors ($\sim$10 MHz). Averaging over the four measurements gives a common line width of 107$\pm$20 km s$^{-1}$.

\begin{table*}
\centering  
\caption{Line parameters for CO(10-9) and CO(11-10). The frequencies are observed values and the continuum contributions are obtained from Figure \ref{fig9}. The FWHM and peak values are obtained from Gaussian fits to the observed line profiles. The references in square brackets are for [1]: \citet{Riechers2011d}; [2]: \citet{Tuananh2013}, \citet{Tuananh2014}; [3]: this work; [4]: \citet{Hainline2004}}
\label{Table4}        
\begin{tabular}{|c|c|c|c|c|c|}
\hline 
Line & Frequency (GHz) & Continuum (mJy) & FWHM (MHz/km\,s$^{-1}$) & Peak (mJy) & Velocity integrated flux (Jy\,km s$^{-1}$) \\
\hline 
CO(1-0) [1] & 30.367 & $-$ & 11.2$\pm$1.9/111$\pm$19 & 1.8$\pm$0.2 & 0.21$\pm$0.03 \\
\hline
CO(3-2) [4] & 91.088 & $-$ & $-$/360$\pm$60 & ? & 2.9$\pm$? \\
\hline
CO(7-6) [2] & 212.50 & 4.4 $\pm$0.5 & 85$\pm$10/120 $\pm$14 & 40$\pm$5 & 5.0$\pm$1.0 \\
\hline
CO(10-9) [3] & 303.48 & 22.7 $\pm$ 0.5 & 89$\pm$16/89$\pm$16 & 47$\pm$8 & 4.5$\pm$1.0 \\
\hline
CO(11-10) [3] & 333.80 & 27.5$\pm$0.5 & 113$\pm$30/102$\pm$28 & 24$\pm$5 & 2.5$\pm$0.9\\
\hline
\end{tabular} 
\end{table*}

\section{Summary and discussion}
\subsection{Summary}
The new observations of the quasar host RX J0911 that have been presented here are an important complement to what was already known of this galaxy. It occupies an eccentric situation in the family of quasar hosts at $z\sim$3, with CO luminosity (related to the gas mass), FIR luminosity (related to the dust mass), line width (related to the dynamical mass) and X ray luminosity (related to the black hole mass) typically five times smaller than usual for such galaxies (Figure \ref{fig3}, \citet{Carilli2013}).

The main contribution of the present work is evidence for the gas and dust components to be compact and concentric with the central black hole, the former being \mbox{$\sim$3.4$\pm$0.4} times more extended than the latter. The availability of HST images, showing also the lens galaxy, has made it possible to define the lensing mechanism with particularly good precision in the image plane, $\pm$29 mas in $x$ and $\pm$11 mas in $y$. The concentricity of the three sources, quasar, gas and dust, measured to better than 0.31 kpc for the quasar versus \mbox{358 GHz} continuum emissions and to better than 1.1 kpc for the quasar versus CO(7-6) emissions, argues against any important and recent merger contribution. The compactness of the dust and gas sources, with respective FWHM values of 0.76$\pm$0.08 kpc and 2.6$\pm$0.3 kpc, adds to the argument. 

Another important contribution is the detection of high J excitations of the CO molecule. It is remarkable that they confirm the extreme narrowness of the CO line (\mbox{107$\pm$20 km s$^{-1}$} on average). The measurements of the CO ladder displayed in Figure \ref{fig11} as ratios to $S_{1-0}$ extend beyond the typical ladder predicted by a single component Large Velocity Gradient (LVG) approximation, suggesting that the temperature and/or density of the gas are/is on the high side. The sensitivity and spatial resolution with which the CO(11-10) and CO(10-9) emissions have been observed are insufficient to resolve the gas component and measure its size as could be done for CO(7-6), but their mere detection suggests a possibly significant contribution of the central AGN to gas and dust heating.

In the remaining of this section, we explore the consequences of the present results in the framework of our present knowledge of galaxy evolution at high redshifts.

\subsection{Comparison with high redshift galaxies for which the sizes of both the gas and dust components have been measured}
Of the galaxies listed in Table \ref{Table1}, five are interpreted as hosting major wet mergers (J123707+6214, SPT0538-50, SMMJ02399-0136, SDP.81 and BRI 1335-0417). One, \mbox{PSS J2322+1944}, a $z$=4.12 quasar host with a star formation rate of $\sim$680 solar masses per year, displays a complex image pattern interpreted as resulting from the lensing of a source significantly offset from the central quasar and hosting interaction possibly caused by a major merger.

This leaves only two galaxies with gas and dust distributions comparable with those of RX J0911: the Cloverleaf ($z$=2.56) and APM 08279+5255 ($z$=3.91). The CO emission of the Cloverleaf, with line widths of 300 to 400 \mbox{km s$^{-1}$} FWHM, is interpreted by \citet{Venturini2003} as originating from a rotating disc-like structure with a characteristic radius of 0.8 kpc, concentric with the quasar. \citet{Bradford2009} measure the CO ladder up to J$_\textrm{up}$=9 with no sign of decrease and suggest that the Cloverleaf host is undergoing a massive starburst, but that it has additional energy input into the ISM via hard X-rays originating in the accretion zone. \citet{Riechers2011d} and \citet{Weiss2003} measure CO emission with \mbox{$L'_\textrm{CO}$$\sim18$$\times$$10^9$ K km s$^{-1}$ pc$^2$}. The FIR luminosity is 22$\times10^{11}$ solar luminosities and the dust emission is confined to the same central region as the gas emission. Unfortunately, the lens parameters are difficult to evaluate with precision. \citet{Granato1996} measure an X-ray luminosity of \mbox{3.5$\times$10$^{46}$ erg s$^{-1}$} giving a black hole mass of 6.8$\times$$10^8$ solar masses.

APM 08279+5255, a radio-quiet quasar host, has CO line widths in the 400 to 500 km s$^{-1}$ range and is lensed into a quad in a similar configuration to RX J0911. \citet{Riechers2009a} give a very detailed analysis of the multi-wavelength observations made of this galaxy and argue in favour of a modest lensing magnification, $\sim$4. The size of the CO component is $\sim$1.1 kpc FWHM, the CO and FIR luminosities are respectively \mbox{26$\times$10$^9$ K km s$^{-1}$ pc$^2$} and  240$\times$10$^{11}$ solar luminosities; together with a black hole mass of $\sim$230$\times$10$^8$ solar masses, this makes APM 08279+5255 a dust- and gas-rich galaxy with a very massive, active black hole in its centre. The gas in the central region is accordingly both dense and warm, the contribution of the starburst to the heating of the gas and dust being rather modest: ~10\% as estimated by \citet{Riechers2009a} and 35\% as estimated by \citet{Ferkinhoff2010}. Evidence is obtained for the central black hole growing faster than the stars in the early phase of galaxy formation. 

In addition to the galaxies listed in Table \ref{Table1}, there exist several other high redshift galaxies for which the sizes of the dust and gas components have been measured and found compact. A complete review is beyond the scope of the present article but two examples are worth quoting as an illustration: \mbox{MACSJ0032-arc} and \mbox{SDSS J1148}.  \mbox{MACSJ0032-arc} \mbox{\citep{Dessauges-Zavadsky2016}} is a $z$=3.63 main sequence star forming galaxy lensed into an arc structure by a complex set of lenses with a magnification of order 60. The star formation region, with a rate in excess of 3000 solar masses per year, is concentrated in the centre of the galaxy, with most of the gas and dust residing within in a diameter of $\sim$1.2 kpc. It is surrounded by two UV-bright star-forming regions, possibly revealing a merger. Most of the star forming rate originates from thermal dust emission. CO emission, with line widths in excess of 300 km s$^{-1}$ FWHM, remains high up to J=6 but higher J levels have not been observed. SDSS J114816.64+525150.3 is a $z$=6.42 quasar host with a black hole of $\sim$30$\times$10$^8$ solar masses \citep{Willott2003} in its centre. CO emission, with line widths of \mbox{$\sim$300 km s$^{-1}$} FWHM, and dust emission, with a far infrared luminosity of $\sim$120$\times$10$^{11}$ solar luminosities, are concentric and confined within radii of $\sim$2.5 kpc and 0.75 kpc respectively \citep{Riechers2009b}.

In summary, compactness of the star formation region about the galactic centre is a property shared by galaxies covering a very broad range of scales, with black hole masses up to two orders of magnitude above that of RX J0911. Its main virtue is to exclude major mergers and multiple or eccentric star bursts.

\subsection{Comparison with high redshift galaxies displaying narrow CO lines}
RX J0911 has the narrowest CO line of all high $z$ galaxies known to us. Table \ref{Table5} lists such galaxies having a CO line width not exceeding 200 km s$^{-1}$ \citep{Carilli2013}. The CO line width is expected to result from several possible causes such as rotation of the gas around the centre of the galaxy, multiple sources associated with different starbursts in a same galaxy, merging galaxies, etc. Indeed, most ultra-luminous infrared galaxies and many sub-millimetre galaxies are known to be the seats of wet mergers and/or multiple starbursts. In the case of RX J0911, the evidence for ellipticity of the gas component (3.3 standard deviations away from a circular source hypothesis) and for velocity gradient along the major axis (25 km s$^{-1}$ kpc$^{-1}$, 4.5 standard deviations from zero) make it unlikely that the gas reservoir be a rotating disc seen close to face on. Moreover, the limits obtained in the present work on the size of the gas and dust components (respectively 2.6$\pm$0.3 and 0.76$\pm$0.08 kpc FWHM) and their concentricity with respect to the quasar make multiple star bursts and/or wet mergers equally unlikely.

\begin{table*}
\centering 
\caption{Galaxies having redshift in excess of 2 and a CO line width not exceeding 200 km s$^{-1}$. The type (column 3) is abbreviated as QSO for quasars, SMM for sub-millimetre galaxies and LBG for Lyman-break galaxies. The FWHM of the CO line (column 5) is averaged over all observed lines. ${L_\textrm{FIR}}$, ${L'_\textrm{CO}}$ and $\textrm{SFR}$ and M$_\textrm{BH}$ (columns 6 to 8) are respectively the far-infrared luminosity in units of $10^{11}$ solar luminosities, the CO luminosity in units of $10^9$ K km s$^{-1}$ pc$^2$ and the star formation rate in units of solar masses per year. Column 9 gives references (a to p) and comments (1 to 3). References stand for: a) \citet{Simpson2012}; b) \citet{Walter2011}; c) \citet{Weiss2005}; d) \citet{Downes2003}; e) \citet{Harris2010}; f) \citet{Baker2004}; g) \citet{Riechers2010}; h) \citet{Riechers2011d}; i) \citet{Ao2008}; j) \citet{Vandenbout2004}; k) \citet{Weiss2005}; l) \citet{Coppin2007}; m) \citet{Lestrade2010} and \citet{Lestrade2011}; n) \citet{Coppin2010}; o) \citet{Combes2012}; p) \citet{Wang2010}. Comments stand for: 1) estimates a gas component size of 700 pc from indirect luminosity arguments; 2) assuming a magnification of 100; 3) the source gives evidence for a merger of two galaxies, one of which only displays a narrow CO line.}
\label{Table5}        
\begin{tabular}{|c|c|c|c|c|c|c|c|c|}
\hline 
Source & $z$ & Type & J$_\textrm{up}$ & FWHM & $L_\textrm{FIR}$ & $L'_\textrm{CO}$ & SFR & R/C \\
\hline 
IRAS F10214+4724   & 2.29  & QSO  & 1,3,4,6,7  & 215$\pm$7  &  35  &  5.8  &  -  & hijk   \\
\hline
J0908-0034   & 2.55  & QSO  & 3  & 125$\pm$25  & 31   &  7.7  &  -  &  a  \\
\hline
SMM J14011+0252   & 2.57  & QSO  & 1,2,3,7  & 192$\pm$9  &  2.6  &  4  &  -  &  bcde;1  \\
\hline
MS1512-cB58   & 2.73  & LBG  & 3  & 174$\pm$43  & 0.8   & 0.6   &  24  &  fg  \\
\hline
RX J0911.4+0551   & 2.80  & QSO  & 1,7,10,11  & 107$\pm$20  & 15   &  4  & 230   &  -  \\
\hline
J213512.73-010143   & 3.07  & LBG  & 3  & 190$\pm$24  & 3.4   &  3  &  60  &  lg  \\
\hline
MM18423+5938   & 3.93  & SMM  & 1,2,4,6,7  & 188$\pm$10  & 48   &  2.7  & 830   & m;2   \\
\hline
J033229.4-275619   & 4.76  & QSO  & 2  & 160$\pm$65  &  60  &  20  &  1000  &  n  \\
\hline
J091828.6+514223   & 5.24  & SMM  & 2,5,6,7  & 145$\pm$15  &  100  &  20  &  1600  &  o;3  \\
\hline
J 1044-0125   & 5.78  & QSO  & 6  & 160$\pm$50  &  52  &  8  & 530   &  p  \\
\hline

\end{tabular} 
\end{table*}

\begin{figure*}
\centering
\includegraphics[height=5.cm,trim=0.cm 0.cm 0.cm 0.cm,clip]{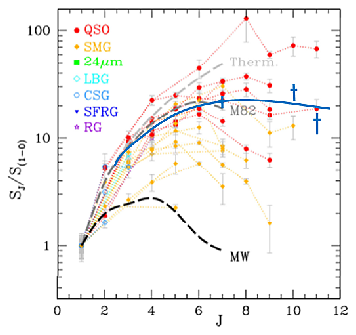}
\includegraphics[height=5.cm,trim=0.cm 0.cm 0.cm 0.cm,clip]{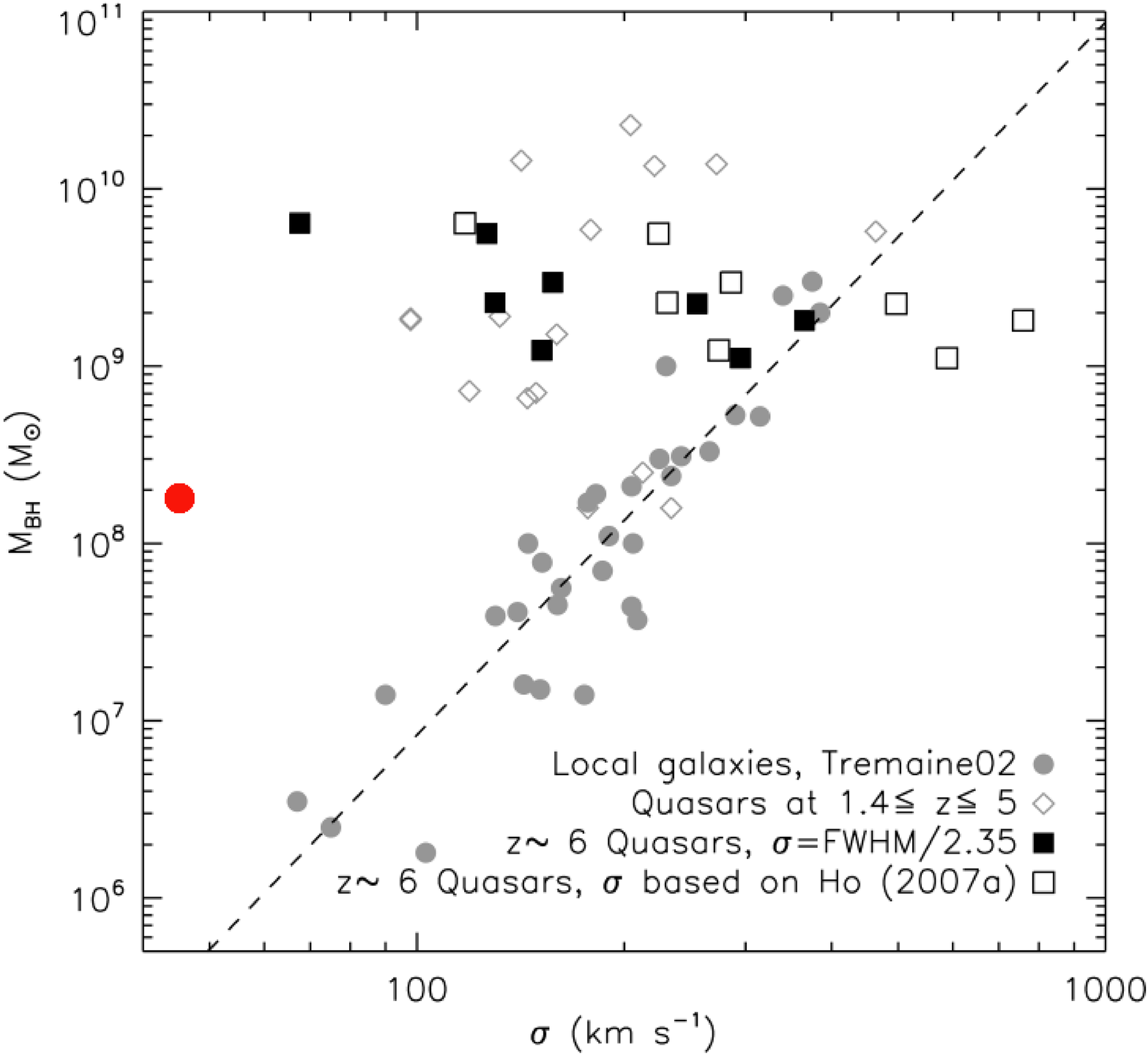}
\includegraphics[height=5.5cm,trim=0.cm 0.cm 0.75cm 0.cm,clip]{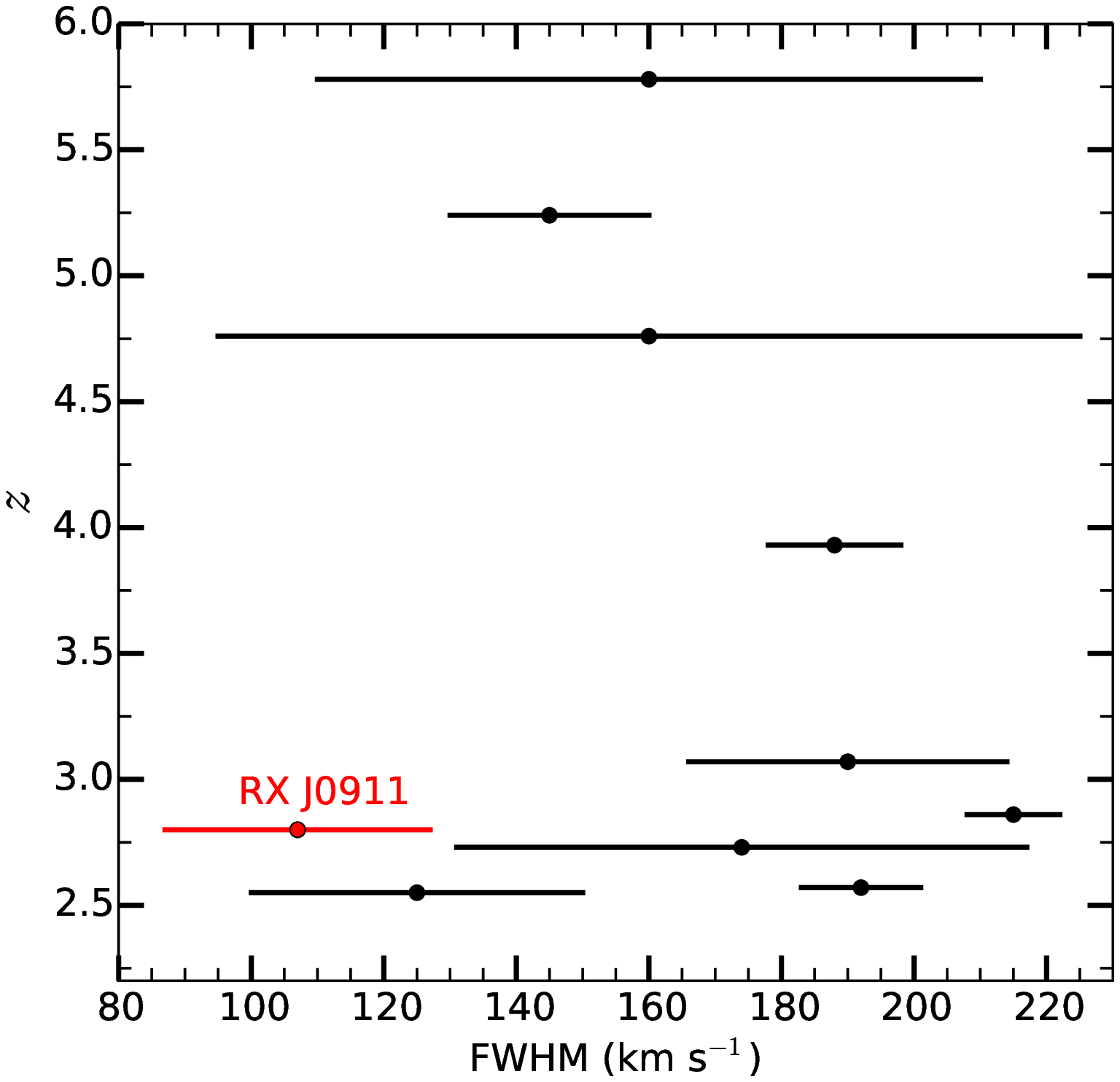}
\caption{Left: CO emission ladder normalised to CO(1-0) for rotational states as a function of the angular momentum $J$ of the initial state \citep[from][]{Carilli2013}. Measured fluxes are displayed for different kinds of galaxies separately as indicated in the insert.  The results of the present work are shown as blue crosses. The blue line is drawn to guide the eye. Middle: Black hole masses (M$_\textrm{BH}$) vs bulge velocity dispersion ($\sigma$) \citep[from][]{Wang2010}. The dashed line and filled circles denote the local M$_\textrm{BH}$-$\sigma$ relationship \mbox{\citep{Tremaine2002}}. The open diamonds are for high redshift quasars with 1.4<$z$<5. The squares are for $z$$\sim$6 quasars \mbox{\citep{Wang2010}}. The red filled circle is for RX J0911. Right: CO line width (FWHM) vs. redshift for the galaxies listed in Table \ref{Table5}.}
\label{fig11}
\end{figure*}

To the extent that the CO line width is directly related to the star velocity dispersion ($\sigma^*$), it is expected to be strongly correlated with the mass of the central black hole (M$_\textrm{BH}$). Such a tight correlation is indeed found for local quasars \citep[for a recent review, see][]{Sheinis2016}. However, this is no longer the case for high redshift quasars \citep{Wang2010} as illustrated in Figure \ref{fig11} (middle panel). Most of the high-$z$ CO-detected quasars are above the local M$_\textrm{BH}$-$\sigma^*$ relation with offsets exceeding one order of magnitude in black hole mass \citep{Shields2006}. Only a small part of these offsets can be blamed on the inclination of the disc on the sky plane. This reflects the observation of large mass ratios between the black hole and the bulge, an effect of the early and more rapid formation of the central supermassive black hole with respect to the bulge. In the case of RX J0911, the local relation predicts a black hole mass nearly 300 times too small or a CO line width nearly 5 times too large, depending on what one starts from. 

In this context, it is interesting to compare RX J0911 with the high redshift galaxies listed in Table \ref{Table5} and illustrated in Figure \ref{fig11} (right panel). RX J0911 is the only galaxy for which a measurement of the size of the gas and/or dust component is available but a common feature observed in most of these galaxies is the absence of an extended massive gas reservoir; in most cases, their properties are qualitatively similar to those of the lower-$z$ sub-millimetre galaxies studied by \citet{Greve2005}.   

As many high redshift luminous galaxies display evidence for mergers and/or multiple starbursts, these are likely to contribute a significant part of the CO line width; as a consequence, the observation of a narrow line width is evidence against such activity, irrespective of the inclination of the gas disc with respect to the sky plane. In the present case of RX J0911, the evidence for a velocity gradient along the elongation of the gas component suggests that most, if not all, of the CO line width is likely to be associated with gas rotation. 

In the cases where the CO ladder has been measured, RX J0911 is the only galaxy listed in Table \ref{Table5} for which high J lines have been detected. Earlier observations have suggested that in the others the main source of gas heating is from star formation rather than from the central AGN, but one must underline that before the recent availability of band 7 detection at ALMA, searching for such high excitations was simply out of reach. In the case of RX J0911, the strong concentration of the star formation region around the central black hole makes it difficult to disentangle the respective contributions of stars and AGN to gas heating and a significant contribution of the latter cannot be excluded.

\subsection{Comparison with high redshift galaxies displaying high level excitation of the CO molecules}

Table \ref{Table6} lists high redshift galaxies for which CO excitations of J$_\textrm{up}$ of 9 or above have been detected. Other galaxies, such as J1148 \citep{Riechers2013} and BR 1202 \mbox{\citep{Salome2012}}, which remain at a high level of excitation up to J$_\textrm{up}$=8, may be of the same category but higher J excitations have not been searched for.

CO ladders are commonly interpreted in the framework of the Large Velocity Gradient approximation \mbox{\citep{Greve2014}}, based on the existence of important turbulence allowing photons to escape in spite of optical thickness. Temperatures of $\sim$40 K are obtained for typical high redshift galaxies \mbox{\citep{Carilli2013, Weiss2007a}} assuming molecular hydrogen densities of $\sim$10$^3$ cm$^{-3}$. Higher J excitations can be accounted for by a temperature typically 20 K larger with molecular hydrogen densities of $\sim$10$^4$ cm$^{-3}$. However, different excitations probe different temperatures and therefore different regions: a two component description is likely to be more appropriate in cases where the central AGN contributes significant heating.

High J excitations have also been used as probes of the gas heating mechanism in local starbursts and/or mergers, such as NGC 6240 and Arp 193 \mbox{\citep{Papadopoulos2014}} and found to peak at moderate values of J$_{\textrm{up}}$. On the contrary, starburst galaxies M82 \citep{Panuzzo2010} and NGC 253 \citep{Rosenberg2014} display a CO ladder that stays high up to J$_{\textrm{up}}$=13, requiring an additional heating source, most likely mechanical, associated with shocks and/or turbulences. 

Both sub-millimetre galaxies listed in Table \ref{Table6}, \mbox{J213511} and \mbox{HLSW-01}, have CO ladders peaking at J$_\textrm{up}$ $\sim$ 6 to 7. The four other galaxies (mostly QSOs), including RX J0911, display excitations that remain high up to the larger values of J$_\textrm{up}$. The Cloverleaf and APM 08279 have been discussed in Section 8.2. In the latter case heating from the central AGN is held responsible for the higher J excitation. In the case of the Cloverleaf, both sources seem to contribute, however with dominance from the starburst \mbox{\citep{Riechers2011f}}. SPT 21323 is a $z$=4.77 lensed hyper-luminous infrared galaxy with a star formation rate of 1120$\pm$200 solar masses per year. The excitation remains high at J$_\textrm{up}$=12, the very short depletion timescale (34$\pm$13 Myr) indicates that this source is an extreme starburst with moderate gas content, but very high star formation efficiency.

Here again, as in Section 8.2, high J$_\textrm{up}$ CO excitation is a property shared by galaxies covering a very broad range of scales, with black hole masses up to two orders of magnitude above that of RX J0911. It reveals the importance of central heating, either directly from the AGN or from an intense star burst surrounding it. 

\begin{table}
\centering 
\caption{High redshift galaxies for which CO excitations of J$_\textrm{up}$ of 9 or above have been detected. The type (column 3) is abbreviated as QSO for quasars, SMM for submillimetre galaxies and ULIRG for Ultrahigh luminosity infrared galaxies. L$_\textrm{FIR}$ (column 5) is the far-infrared luminosity in units of 10$^{11}$ solar luminosities. Column 6 gives references (1 to 7): 1) \citet{Danielson2011}; 2) \citet{Bradford2009}; 3) \citet{Scott2011}; 4) \citet{Bradford2011}; 5) \citet{Downes1999}; 6) \citet{Weiss2007b}; 7) \citet{Bethermin2016}}
\label{Table6}        
\begin{tabular}{|c|c|c|c|c|c|}
\hline 
Source & $z$ & Type & J$_\textrm{up}$ & $L_\textrm{FIR}$ & Ref. \\
\hline
J213511    &  2.32  &  SMM  &   1,3,4,5,6,7,8,9  &  23  & 1 \\
\hline
Cloverleaf    & 2.56   &  QSO  &  1,3,4,5,6,7,8,9  &  22  & 2  \\
\hline
RX J0911    &  2.80  &  QSO  &  1,7,10,11  &  15  &  - \\
\hline
HLSW-01    &  2.96  &  SMM  &  1,3,5,7,9,10  &  143  &  3 \\
\hline
APM08279    &  3.91  & QSO   &  1,2,4,6,8,9,10,11  & 240   &  4,5,6\\
\hline
SPT 21323    &  4.77  &  ULIRG  &  2,5,12  &  68  &  7\\
\hline

\end{tabular} 
\end{table}

\subsection{Conclusion}
According to current understanding, galaxies first formed from the gravitational collapse of gas and dark matter inhomogeneity, with rapid production of massive short-lived stars resulting in coeval and prompt generation of the central black hole and enrichment of the ISM in metallic species. Subsequent growth occurred via both mergers and additional gas accretion from the intergalactic medium, with star formation rates reaching a maximum at redshifts of order $\sim$3, when the constantly decreasing ratio of gas to star mass crossed unity. How does RX J0911 fit into such a picture? Its main characteristics are the compactness of the dust component, the narrow width of the CO emission lines, the extension of the CO ladder to high J values and overall scaled-down values of relevant masses and luminosities in comparison with quasar hosts at similar redshifts. The most natural interpretation is that of a young galaxy in an early stage of its evolution, having experienced no recent major mergers, star formation being concentrated in its centre. When compared with other high redshift star forming galaxies, the Cloverleaf may be that closest to resemble it, although scaled-up by a factor of order 4. Additional observations are necessary to deepen our understanding of the physics governing the evolution of RX J0911. In particular, the high spatial resolution available from ALMA long baselines should be exploited to map the emission of CO lines for at least two representative values of J$_\textrm{up}$, such as 5 and 10, as well as the emission of atomic lines, such as [C$_\textrm{II}$] and of high dipole moment molecules.

\section*{Acknowledgements}
This paper makes use of the following ALMA data: ADS/JAO.ALMA\#2011.0.00307.S. ALMA is a partnership of ESO (representing its member states), NSF (USA) and NINS (Japan), together with NRC (Canada), NSC and ASIAA (Taiwan), and KASI (Republic of Korea), in cooperation with the Republic of Chile. The Joint ALMA Observatory is operated by ESO, AUI/NRAO and NAOJ. The data are retrieved from the JVO portal (http://jvo.nao.ac.jp/portal) operated by the NAOJ. We are indebted and very grateful to the ALMA partnership, who are making their data available to the public after a one year period of exclusive property, an initiative that means invaluable support and encouragement for Vietnamese astrophysics. We particularly acknowledge friendly and patient support from the staff of the ALMA Helpdesk. We are grateful to Pr Fran\c{c}oise Combes and Pr St\'{e}phane Guilloteau who clarified for us some points related with astrometry. We thank Chelsea Sharon for having brought a recent work of her to our attention. We thank the anonymous referee for very useful comments that helped to significantly improve the presentation of our work. Financial support is acknowledged from the Vietnam National Satellite Centre (VNSC/VAST), the Vietnam National Foundation for Science and Technology Development (NAFOSTED) funding agency under contract 103.99-2015.39, the World Laboratory, the Odon Vallet Foundation and the Rencontres du Viet Nam.





\begin{thebibliography}{99}

\bibitem[\protect\citeauthoryear{Alexander et al.}{2005a}]{Alexander2005a}
 Alexander D.M., Smail I., Bauer F.E. et al. 2005a, Nature, 434, 738
\bibitem[\protect\citeauthoryear{Alexander et al.}{2005b}]{Alexander2005b}
Alexander D.M., Bauer F.E., Chapman S.C. et al. 2005b, ApJ, 632, 736
\bibitem[\protect\citeauthoryear{ALMA Partnership et al.}{2015}]{ALMA2015}
  ALMA Partnership et al. 2015, ApJL, 808, L4, arXiv:1503.02652
\bibitem[\protect\citeauthoryear{Tuan-Anh et al.}{2013}]{Tuananh2013}
  Anh P.T., Boone F. , Hoai D.T. et al., 2013, A\&A, 552, L12
\bibitem[\protect\citeauthoryear{Tuan-Anh}{2014}]{Tuananh2014}
Anh P.T. 2014, {\it Observations millim\'etriques/submillim\'etriques de galaxies lentill\'ees gravitationellement \`a haut redshift}, PhD thesis, U. Toulouse 3 Paul Sabatier, Toulouse, France and Ha Noi IOP, Viet Nam  
\bibitem[\protect\citeauthoryear{Ao et al.}{2008}]{Ao2008}
Ao Y., Wei{\ss} A., Downes D. et al. 2008, A\&A, 491, 747
\bibitem[\protect\citeauthoryear{Aravena et al.}{2016}]{Aravena2016}
Aravena M., Spilker J.S., Bethermin M. et al. 2016, MNRAS, 457, 4406A
\bibitem[\protect\citeauthoryear{Bade et al.}{1995}]{Bade1995}
Bade N., Fink H.H., Engels D. et al. 1995, A\&AS, 110, 469
\bibitem[\protect\citeauthoryear{Bade et al.}{1997}]{Bade1997}
Bade N., Siebert J., Lopez S. et al. 1997, A\&AS, 317 L13 
\bibitem[\protect\citeauthoryear{Baker et al.}{2004}]{Baker2004}
Baker A.J., Tacconi L.J., Genzel R. et al., {\it The Neutral ISM in Starburst Galaxies ASP Conference Series}, Vol. 320, 2004, S. Aalto, S. Huttemeister \& A. Pedlar ed.
\bibitem[\protect\citeauthoryear{Barvainis \& Ivison}{2002}]{Barvainis2002}
Barvainis R. \& Ivison R., 2002, ApJ, 571, 712
\bibitem[\protect\citeauthoryear{Bethermin et al.}{2016}]{Bethermin2016}
B\'ethermin M., De Breuck C., Gullberg B. et al. 2016, A\&A, 586, L7
\bibitem[\protect\citeauthoryear{Bothwell et al.}{2013}]{Bothwell2013}
Bothwell M.S., Smail I., Chapman S.C. et al. 2013, MNRAS, 429, 3047
\bibitem[\protect\citeauthoryear{Bradford et al.}{2009}]{Bradford2009}
Bradford C.M., Aguirre J.E., Aikin R. et al. 2009, ApJ, 705, 112
\bibitem[\protect\citeauthoryear{Bradford et al.}{2011}]{Bradford2011}
Bradford C., Bolatto A.D., Maloney P. et al. 2011, ApJL, 741, L37
\bibitem[\protect\citeauthoryear{Burud et al.}{1998a}]{Burud1998a}
Burud I., Courbin F., Lidman C. et al. 1998a, ApJ, 501L, 5B
\bibitem[\protect\citeauthoryear{Burud et al.}{1998b}]{Burud1998b}
Burud I., Courbin F., Lidman C. et al. 1998b, Mnsgr, 92, 29B
\bibitem[\protect\citeauthoryear{Carilli et al.}{2001}]{Carilli2001}
Carilli C.L., Bertoldi F., Omont A., Cox P., McMahon R.G., \& Isaak K. 2001, AJ, 122, 1679
\bibitem[\protect\citeauthoryear{Carilli et al.}{2003}]{Carilli2003}
 Carilli C.L., Lewis G.F., Djorgovski S.G. et al. 2003, Science 1082600, arXiv:astro.ph/0304124
\bibitem[\protect\citeauthoryear{Carilli \& Walter}{2013}]{Carilli2013}
  Carilli C.L. \& Walter F. 2013 ARA \& A, 51/1, 105
\bibitem[\protect\citeauthoryear{Combes et al.}{2012}]{Combes2012}
Combes F., Rex M., Rawle T.D. et al. 2012, A\&A, 538L, 4C
\bibitem[\protect\citeauthoryear{Coppin et al.}{2007}]{Coppin2007}
Coppin K.E.K., Swinbank A.M., Neri, R. et al. 2007, ApJ, 665, 936
\bibitem[\protect\citeauthoryear{Coppin et al.}{2010}]{Coppin2010}
Coppin K.E.K., Chapman S.C., Smail I. et al. 2010, MNRAS, 407L, 103  
\bibitem[\protect\citeauthoryear{Danielson et al.}{2011}]{Danielson2011}
Danielson A.L.R., Swinbank A.M., Smail I. et al. 2011, MNRAS, 410, 1687
\bibitem[\protect\citeauthoryear{Dessauges-Zavadsky et al.}{2016}]{Dessauges-Zavadsky2016}
Dessauges-Zavadsky M., Zamojski M., Rujopakarn W. et al. 2016, arXiv:1610.08065
\bibitem[\protect\citeauthoryear{Downes et al.}{1999}]{Downes1999}
  Downes D., Neri R., Wiklind. T. et al. 1999, ApJ, 513, L1
\bibitem[\protect\citeauthoryear{Downes \& Solomon}{2003}]{Downes2003}
Downes D. \& Solomon P.M. 2003, ApJ, 582,37
\bibitem[\protect\citeauthoryear{Fan et al.}{2009}]{Fan2009}
Fan, L.L., Wang, H.Y., Wang, T. et al. 2009, ApJ, 690, 1006
\bibitem[\protect\citeauthoryear{Ferkinhoff et al.}{2010}]{Ferkinhoff2010}
Ferkinhoff C., Hailey-Dunsheath S., Nikola T. et al. 2010, ApJL, 714, L147
\bibitem[\protect\citeauthoryear{Ferkinhoff et al.}{2015}]{Ferkinhoff2015}
  Ferkinhoff C., Brisbin D., Nikola T. et al., 2015, ApJ, 806, 260
\bibitem[\protect\citeauthoryear{Glikman et al.}{2016}]{Glikman2016}
Glikman E., Simmons B., Mailly M. et al. 2016, arXiv:1504.02111
\bibitem[\protect\citeauthoryear{Granato et al.}{1996}]{Granato1996}
Granato G.L., Danese L. \& Franceschini A. 1996, ApJ, 460, L11
\bibitem[\protect\citeauthoryear{Greve et al.}{2005}]{Greve2005}
  Greve T.R., Bertoldi F., Smail I. et al. 2005, MNRAS, 359, 1165
\bibitem[\protect\citeauthoryear{Greve et al.}{2014}]{Greve2014}
  Greve T.R., Leonidaki I., Xilouris E.M. et al. 2014, ApJ, 794, 142
\bibitem[\protect\citeauthoryear{Hagen et al.}{1995}]{Hagen1995}
  Hagen H.J., Groote D., Engels D. et al. 1995, A\&AS 111, 195
\bibitem[\protect\citeauthoryear{Hainline et al.}{2004}]{Hainline2004}
Hainline L.J., Scoville N.Z., Yuin M.S. et al. 2004, ApJ, 609,61
\bibitem[\protect\citeauthoryear{Harris et al.}{2010}]{Harris2010}
  Harris A.I., Baker A.J., Zonak S.J. et al. 2010, ApJ, 723, 1139 
\bibitem[\protect\citeauthoryear{Hezaveh et al.}{2013}]{Hezaveh2013}
Hezaveh Y. D., Marrone D. P., Fassnacht C. D. et al. 2013, ApJ, 767, 132
\bibitem[\protect\citeauthoryear{Hjorth et al.}{2002}]{Hjorth2002}
Hjorth J., Burud I., Jaunsen A.O. et al. 2002 ApJ, 572, L11
\bibitem[\protect\citeauthoryear{Hoai et al.}{2013}]{Hoai2013}
Hoai D.T., Nhung P.T.T., Tuan-Anh P. et al. 2013, RAA, 13/7, 803
\bibitem[\protect\citeauthoryear{Ivison et al.}{2010}]{Ivison2010}
Ivison R.J., Smail I., Papadopoulos P.P. et al. 2010, MNRAS, 404, 198 
\bibitem[\protect\citeauthoryear{Kneib et al.}{2000}]{Kneib2000}
Kneib J.-P., Cohen J.G. and Hjorth J. 2000, ApJ, 544, L35
\bibitem[\protect\citeauthoryear{Kochanek}{2002}]{Kochanek2002}
  Kochanek C.S. 2002, Gravitational Lenses, {\it the Distance Ladder and the Hubble Constant: a New Dark Matter Problem}, arXiv:astro-ph 0204043v1
\bibitem[\protect\citeauthoryear{Lestrade et al.}{2010}]{Lestrade2010}
  Lestrade J.F., Combes F.,  Salom\'e P. et al. 2010, A\&A, 522, L4
\bibitem[\protect\citeauthoryear{}{2011}]{Lestrade2011}
Lestrade J.F., Carilli C.L.,  Thanjavur K. et al. 2011, ApJL, 739, 30L 
\bibitem[\protect\citeauthoryear{McLure \& Dunlop}{2004}]{McLure2004}
McLure R.J. \& Dunlop J.S. 2004, MNRAS, 352, 1390
\bibitem[\protect\citeauthoryear{Momjian et al.}{2007}]{Momjian2007}
Momjian E., Carilli C.L., Riechers D.A. et al. 2007, AJ, 134, 694, arXiv:0706.0899
\bibitem[\protect\citeauthoryear{Morrison et al.}{2010}]{Morrison2010}
  Morrison G.E., Owen F.N., Dickinson M. et al., 2010, ApJS, 188, 178
\bibitem[\protect\citeauthoryear{Panuzzo et al.}{2010}]{Panuzzo2010}
Panuzzo P., Rangwala N., Rykala A. et al. 2010, A\&A, 518, L37
\bibitem[\protect\citeauthoryear{Papadopoulos et al.}{2014}]{Papadopoulos2014}
Papadopoulos P.P., Zhang Z.-Y., Xilouris E.M. et al. 2014, ApJ, 788, 153
\bibitem[\protect\citeauthoryear{Richard et al.}{2009}]{Richard2009}
Richard J., Kneib J. Edge A. \& Jullo E. (2009), quoted by Ivison et al. (2010)
\bibitem[\protect\citeauthoryear{Riechers et al.}{2008a}]{Riechers2008a}
Riechers D.A., Walter F., Brewer B.J. et al., 2008a, ApJ 686, 851
\bibitem[\protect\citeauthoryear{Riechers et al.}{2008b}]{Riechers2008b}
Riechers D.A., Walter F., Carilli C.L. et al. 2008b, AJ 686, L9, arXiv:0808.3774

\bibitem[\protect\citeauthoryear{Riechers et al.}{2009a}]{Riechers2009a}
Riechers D.A., Walter F., Carilli C.L. \& Lewis, G.F. 2009a, ApJ, 690, 463
\bibitem[\protect\citeauthoryear{Riechers et al.}{2009b}]{Riechers2009b}
Riechers D.A., Walter F., Bertoldi F. et al. 2009b, ApJ, 703, 1338
\bibitem[\protect\citeauthoryear{Riechers et al.}{2010}]{Riechers2010}
Riechers D.A., Carilli C.L., Walter F. \& Momjian E. 2010, ApJL, 724, L153
\bibitem[\protect\citeauthoryear{Riechers}{2011a}]{Riechers2011a}
Riechers D.A., {\it{Galaxy Evolution: Infrared to Millimeter Wavelength Perspective, ASP Conference Series}} 2011a, Astronomical Society of the Pacific, arXiv 1103-3897
\bibitem[\protect\citeauthoryear{Riechers}{2011b}]{Riechers2011b}
  Riechers D.A., 2011b, ApJ, 730, 108
\bibitem[\protect\citeauthoryear{Riechers et al.}{2011c}]{Riechers2011c}
Riechers D.A., Hodge J., Walter F. et al., 2011c, ApJ, 739L, 31R
\bibitem[\protect\citeauthoryear{Riechers et al.}{2011d}]{Riechers2011d}
  Riechers D.A., Carilli C.L., Maddalena R.J. et al., 2011d, ApJ, 739L, 32R
\bibitem[\protect\citeauthoryear{Riechers et al.}{2011e}]{Riechers2011e}
Riechers D.A., Carilli L.C., Walter F. et al., 2011e, ApJ, 733L, 11R
  \bibitem[\protect\citeauthoryear{Riechers et al.}{2011f}]{Riechers2011f}
Riechers D.A., Walter F., Carilli C.L. et al. 2011f, ApJ, 726, 50
  \bibitem[\protect\citeauthoryear{Riechers et al.}{2013}]{Riechers2013}
Riechers, D.A., Bradford C.M., Clements D.L. et al. 2013, Nature, 496, 329

\bibitem[\protect\citeauthoryear{Rosenberg et al.}{2014}]{Rosenberg2014}
Rosenberg M.J.F., Kazandjian M.V., van der werf P.P. et al. 2014, A\&A, 564, A126
\bibitem[\protect\citeauthoryear{Rybak et al.}{2015a}]{Rybak2015a}
Rybak M., Vegetti S., McKean J.P. et al. 2015a, MNRAS, 451L, 40R
\bibitem[\protect\citeauthoryear{Rybak et al.}{2015b}]{Rybak2015b}
Rybak M., Vegetti S., McKean J.P. et al. 2015b, MNRAS, 453, L26
\bibitem[\protect\citeauthoryear{Saha \& Williams}{2003}]{Saha2003}
  Saha P. \& Williams L.L.R., 2003, AJ, 125, 2769
\bibitem[\protect\citeauthoryear{Salome et al.}{2012}]{Salome2012}
  Salom\'e P., Gu\'elin M., Downes D. et al. 2012, A\&A, 545, 57
\bibitem[\protect\citeauthoryear{Scott et al.}{2011}]{Scott2011}
  Scott K., Lupu R., Aguirre J. et al. 2011, ApJ, 733, 29
\bibitem[\protect\citeauthoryear{Sharon et al.}{2016}]{Sharon2016}
  Sharon C.E., Riechers D.A., Hodge J. and al. 2016, ApJ, 827, 18S
\bibitem[\protect\citeauthoryear{Sheinis et al.}{2016}]{Sheinis2016}
  Sheinis A.I. \& Lopez-Sanchez A.R. 2016, submitted to AJ, {\it{Quasar Host Galaxies and the MS MBH-$\sigma^*$ Relation}}
\bibitem[\protect\citeauthoryear{Shields et al.}{2006}]{Shields2006}
  Shields G.A., Menezes K.L., Massart C.A. \& Vanden Bout P. 2006, ApJ, 641, 683
\bibitem[\protect\citeauthoryear{Simpson et al.}{2012}]{Simpson2012}
  Simpson J.M., Smail I., Swinbank A.M. et al. 2012, MNRAS, 426, 3201
\bibitem[\protect\citeauthoryear{Spilker et al.}{2015}]{Spilker2015}
  Spilker J.S., Aravena M., Marrone D.P. et al. 2015, ApJ., 811, 124
\bibitem[\protect\citeauthoryear{Tremaine et al.}{2002}]{Tremaine2002}
  Tremaine S., Gebhardt K., Bender R. et al. 2002, ApJ, 574, 740
\bibitem[\protect\citeauthoryear{Ueda}{2015}]{Ueda2015}
  Ueda Y. 2015, Proc. Japan Acad. Ser. B Phys. Biol. Sci., 91/5, 175
\bibitem[\protect\citeauthoryear{Vanden Bout et al.}{2004}]{Vandenbout2004}  
Vanden Bout, P. A., Solomon, P. M. and Maddalena, R. J. 2004, ApJ, 614, L97
\bibitem[\protect\citeauthoryear{Venturini \& Solomon}{2003}]{Venturini2003}
  Venturini S. \& Solomon P.M. 2003, ApJ, 590, 740
\bibitem[\protect\citeauthoryear{Walter et al.}{2011}]{Walter2011}
  Walter F., Wei{\ss} A., Downes D. et al. ApJ 2011, 730, 18
\bibitem[\protect\citeauthoryear{Wang et al.}{2010}]{Wang2010}
  Wang R., Carilli C.L., Neri R. et al. 2010, ApJ, 714, 699
\bibitem[\protect\citeauthoryear{Wei{\ss} et al.}{2003}]{Weiss2003}
  Wei{\ss} A., Henkel C., Downes D. and Walter F. 2003, A\&A, 409, L41
\bibitem[\protect\citeauthoryear{Weiss et al.}{2005}]{Weiss2005}
Weiss A., Downes D., Henkel C. \& Walter F. 2005, {\it Starbursts: From 30 Doradus to Lyman Break Galaxies}, R. de Grijs and R.M. Gonz\'alez Delgado ed., Ap. \& Space Sc. Lib., 329, P84
\bibitem[\protect\citeauthoryear{Weiss et al.}{2007a}]{Weiss2007a}
Weiss A., Downes D., Walter F. and Henkel C. 2007a, {\it From Z-Machines to ALMA: (Sub)millimeter Spectroscopy of Galaxies}, ASP Conf. Series, 375
\bibitem[\protect\citeauthoryear{Wei{\ss} et al.}{2007b}]{Weiss2007b}
  Wei{\ss} A., Downes D., Neri R. et al. 2007b, A\&A, 467, 955
\bibitem[\protect\citeauthoryear{Weiss et al.}{2012}]{Weiss2012}
  Weiss A., Walter F., Downes D. et al., 2012, ApJ, 753, 102
\bibitem[\protect\citeauthoryear{Willott et al.}{2003}]{Willott2003}
  Willott C.J., McLure R.J. and Jarvis M.J. 2003, ApJ, 587, L15
\bibitem[\protect\citeauthoryear{Wu et al.}{2009}]{Wu2009}
Wu, J., Vanden Bout, P. A., Evans, II, N. J., \& Dunham, M. M. 2009, ApJ, 707, 988

\end{thebibliography}




\appendix
\section{Astrometry}
Because of the importance of revealing possible offsets between the locations in the sky plane of the line source, the continuum source and the quasar, astrometry needs to be kept under good control. Our approach is as follows:

1) use the HST observations to locate the lensing galaxy (more precisely the lensing potential 1 of \citet{Tuananh2014}, hereafter referred to as the potential) and the quasar source in HST coordinates. The relevant parameters are listed in Table \ref{TableA1}; the lensing potential
includes the elliptical main lensing galaxy G with amplitude, ellipticity and position angles measured by parameters $r_0$, $\epsilon$ and $\varphi_1$ and an external shear of strength $\gamma_0$ and position angle $\varphi_0$ meant to mimic the influence of a satellite galaxy G' and of the galaxy cluster of which G and G' are part; with respect to the best fit to the HST images given in \citet{Tuananh2014}, we have shifted the origin of HST coordinates 29 mas north and 3 mas east from the nominal G position, resulting in a significant improvement of the quality of the fit. The rms deviation between the observed image coordinates and those obtained from the fit is now $\pm$22 mas ($\pm$29 mas in $x$ and $\pm$11 mas in $y$).

2) Accept that the HST and PdBI reference frames may be different but assume that the quasar point source and the centre of the CO(7-6) line emission are identical;

3) obtain offsets (offsets 1) between the HST and PdBI reference frames from a best fit to the PdBI line data of a model using for the source a disc having a radius of 160 milliarcseconds and centred on the quasar;

4) repeat the exercise to obtain offsets (offsets 2) between the HST and ALMA reference frames from a best fit to the ALMA continuum data of a model using for the source a disc having a radius of 40 mas and centred on the quasar;

5) check the consistency of the results by adjusting directly offsets (offsets 3) between the ALMA and PdBI reference frames from a fit between the two sets of data (continuum and line). Note that this lacks rigour because it compares images of sources having different sizes.

The results are summarized in Table \ref{TableA2}. Offsets 1 give the coordinates of the origin of the PdBI frame in the HST frame; offsets 2 give the origin of the ALMA frame in the HST frame; offsets 3 give the origin of the PdBI frame in the ALMA frame, which should be equal to the difference between offsets 1 and offsets 2. 
The very good agreement between the direct (offsets 3) and indirect (offsets 4) evaluations of the relative position of the PdBI vs ALMA reference frames, 15 mas in $x$ and 6 mas in $y$, is merely a consistency cross-check, also showing that the precision of the astrometry is not very sensitive to the size of the source. The question to be addressed next is with which precision we know the relative positions of one of the three reference frames (HST, PdBI and ALMA) with respect to another. In principle, both ALMA and PdBI, using auto-correlation and strong quasars for calibration, are expected to be very accurately positioned. We evaluate the precision as the beam dimensions divided by the signal to noise ratio, namely, for $x$ and $y$ respectively, 28 mas and 68 mas for PdBI and 10 mas and 10 mas for ALMA. The relative positioning of the PdBI vs ALMA frames is obtained by adding these numbers in quadrature: $\sim$30 mas in $x$ and $\sim$70 mas in $y$. However, this evaluation is optimistic as it neglects uncertainties associated with the precise knowledge of the baselines. Assuming that these are typically at the level of 5\% of the beam dimensions would increase the positioning errors of the PdBI frame to $\sim$31 mas in $x$ and 76 mas in $y$ and those of the ALMA frame to $\sim$25 mas for each of $x$ and $y$. Then, the relative positioning uncertainty of the PdBI vs ALMA frames increases to $\sim$40 mas in $x$ and $\sim$80 mas in $y$.

For the HST, the positioning is much less well known: when using the same guide stars, one may expect an accuracy of 50 to 100 mas, but only 200 to 500 mas when using different guide stars. Note that the precision with which the lens model reproduces observations, 29 mas in $x$ and 11 mas in $y$ is irrelevant when comparing frames of reference.
\begin{table}
\centering  
\caption{Parameters of the lensing potential $\varphi=r_0r(1+\epsilon \cos2[\varphi-\varphi_1])^{\sfrac{1}{2}}+\sfrac{1}{2}\gamma_0r^2\cos2[\varphi-\varphi_0]$ and source position (at distance $r_s$ from the origin and position angle $\varphi_s$) giving the best fit to the HST images of the quasar point source \citep{Tuananh2014}.}
\label{TableA1} 
\begin{tabular}{|c|c|c|c|c|c|c|} 
  \hline 
  \multicolumn{7}{|c|}{Lens parameters and source position}\\
  \multicolumn{7}{|c|}{(angles in degrees, distances in arcseconds)} \\
  \hline 
  \multicolumn{5}{|c|}{Lens parameters} & \multicolumn{2}{c|}{Source position} \\
  \hline
  $\rho_0$ & $\epsilon$ & $\varphi_1 $ & $\gamma_0$  & $\varphi_0$ & $r_1$ & $\varphi_1$ \\
  \hline
  1.1085 & $-$0.0237 & 65.0 & 0.309 & 7.32 & 0.4468 & 4.18\\
  \hline
\end{tabular} 
\end{table}

\begin{table}
\centering  
\caption{Offsets (milliarcseconds) between the HST, PdBI and ALMA reference frames.}
\label{TableA2}        
\begin{tabular}{|c|c|c|}
\hline 
 & $x$ offset (mas) & $y$ offset (mas)\\
\hline 
1 (PdBI-HST) & $-$148 & $-$255\\
\hline
2 (ALMA-HST) & $-$115 & $-$141\\
\hline
3 (PdBI-ALMA) & $-$18 & $-$120 \\
\hline
4 (1$-$2) & $-$33 & $-$114\\
\hline
\end{tabular} 
\end{table}
Taking all above considerations in due account, we therefore retain for the PdBI vs ALMA offsets (offsets 3) the values $\Delta x \sim -25 \pm$40 mas and $\Delta y \sim -117\pm$80 mas. They are 0.6$\sigma$ and respectively 1.5$\sigma$ away from zero: it is therefore reasonable to accept that our approach is sensible and use these quantities as evaluations of small positioning errors. Using the ALMA frame as reference (because it is the most accurately positioned) reduces offsets 1 and 2 to common mean values of $-$122 mas in $x$ and $-$138 mas in $y$, again compatible with the positioning errors to be expected from the HST. 

\bsp	
\label{lastpage}
\end{document}